\documentclass[12pt,preprint]{aastex}    % for AJ
\newcommand{\Msol}{M$_{\odot}$}
\newcommand{\HI}{H\,{\sc {i}}~}
\newcommand{\MHI}{M$_{\rm HI}$}
\newcommand{\Msold}{M$_{\odot}$\,yr$^{-1}$}
\newcommand{\Vlsr}{V$_{\rm lsr}$}
\newcommand{\Vexp}{V$_{\rm exp}$}

\newcommand{\Teff}{T$_{\rm eff}$}
\newcommand{\kms}{km\,s$^{-1}$}

\slugcomment{submitted to the AJ, May 5, 2006; accepted, Aug. 30, 2006}

\shorttitle{\HI in circumstellar environments}
\shortauthors{G\'erard \& Le~Bertre}

\begin{document}

%% LaTeX will automatically break titles if they run longer than
%% one line. However, you may use \\ to force a line break if
%% you desire.

\title{Circumstellar atomic hydrogen in evolved stars{\footnote{This paper 
is dedicated to the memory of Marie-Odile Mennessier (1943-2004).}}}

%% Use \author, \affil, and the \and command to format
%% author and affiliation information.
%% Note that \email has replaced the old \authoremail command
%% from AASTeX v4.0. You can use \email to mark an email address
%% anywhere in the paper, not just in the front matter.
%% As in the title, use \\ to force line breaks.

\author{E. G\'erard}
\affil{GEPI, UMR\,8111, Observatoire de Paris, 5 place J. Janssen, 
           F-92195 Meudon Cedex, France}
\email{Eric.Gerard@obspm.fr}

\and

\author{T.~Le~Bertre}
\affil{LERMA, UMR 8112, Observatoire de Paris, 61 av. de l'Observatoire,
           F-75014 Paris, France}
\email{Thibaut.LeBertre@obspm.fr\\}

%% Notice that each of these authors has alternate affiliations, which
%% are identified by the \altaffilmark after each name.  Specify alternate
%% affiliation information with \altaffiltext, with one command per each
%% affiliation.

%% Mark off your abstract in the ``abstract'' environment. In the manuscript
%% style, abstract will output a Received/Accepted line after the
%% title and affiliation information. No date will appear since the author
%% does not have this information. The dates will be filled in by the
%% editorial office after submission.

\begin{abstract}
We present new results of a spectroscopic survey of circumstellar \HI in 
the direction 
of evolved stars made with the Nan\c cay Radiotelescope. The \HI line at 21 cm 
has been detected in the circumstellar shells of a variety of evolved stars\,: 
AGB stars, oxygen-rich and carbon-rich, Semi-Regular and Miras, 
and Planetary Nebulae. The emissions are generally spatially resolved, 
i.e. larger than 4$'$, indicating 
shell sizes of the order of 1 pc which opens the possibility to trace the 
history of mass loss over the past $\sim 10^4-10^5$ years. 
The line-profiles are sometimes composite. 
The individual components have generally a quasi-Gaussian shape; 
in particular they seldom show the double-horn profile that would be expected 
from the spatially resolved optically thin emission of a uniformly expanding 
shell. This probably implies that the expansion velocity decreases outwards in 
the external shells (0.1--1 pc) of these evolved stars.

The \HI line-profiles do not necessarily match those of the CO rotational 
lines. Furthermore, the centroid velocities do not always agree with those 
measured in the CO lines and/or the stellar radial velocities. The \HI 
emissions may also be shifted in position with respect to the central stars. 
Without excluding the possibility of asymmetric mass ejection, we suggest 
that these two effects could also be related to a non-isotropic interaction 
with the local interstellar medium. 

\HI was detected in emission towards several sources ($\rho$ Per, 
$\alpha$ Her, $\delta^2$ Lyr, U CMi) that otherwise have 
not been detected in any radio lines. Conversely it was not detected in the 
two oxygen-rich stars with substantial mass-loss rate, NML Tau and WX Psc, 
possibly because these sources are young with hydrogen in molecular 
form, and/or because 
the temperature of the circumstellar \HI gas is very low ($<$ 5 K).
\end{abstract}

%% Keywords should appear after the \end{abstract} command. The uncommented
%% example has been keyed in ApJ style. See the instructions to authors
%% for the journal to which you are submitting your paper to determine
%% what keyword punctuation is appropriate.

%% Authors who wish to have the most important objects in their paper
%% linked in the electronic edition to a data center may do so in the
%% subject header.  Objects should be in the appropriate "individual"
%% headers (e.g. quasars: individual, stars: individual, etc.) with the
%% additional provision that the total number of headers, including each
%% individual object, not exceed six.  The \objectname{} macro, and its
%% alias \object{}, is used to mark each object.  The macro takes the object
%% name as its primary argument.  This name will appear in the paper
%% and serve as the link's anchor in the electronic edition if the name
%% is recognized by the data centers.  The macro also takes an optional
%% argument in parentheses in cases where the data center identification
%% differs from what is to be printed in the paper.


%% From the front matter, we move on to the body of the paper.
%% In the first two sections, notice the use of the natbib \citep
%% and \citet commands to identify citations.  The citations are
%% tied to the reference list via symbolic KEYs. The KEY corresponds
%% to the KEY in the \bibitem in the reference list below. We have
%% chosen the first three characters of the first author's name plus
%% the last two numeral of the year of publication as our KEY for
%% each reference.

\section{INTRODUCTION}

Hydrogen is the most abundant element in the atmospheres of 
Asymptotic Giant Branch (AGB) stars and in their outflows. It may be found 
in the form of atomic hydrogen, H\,{\sc {i}}, or molecular hydrogen, H$_2$, 
besides relatively minor (in abundance) species.
The fraction of hydrogen in each of these two forms is an important factor 
controlling the atmospheric structures of red giants and also the winds which 
develop from their atmospheres (e.g. Lamers \& Cassinelli 1999). 

It is therefore of interest to study the circumstellar environments of 
AGB stars in the low excitation hydrogen lines. 
Unfortunately molecular hydrogen is presently difficult to detect 
because its lowest rotational lines lie in a region of the spectrum which 
cannot be accessed from the ground, and because a high spectral 
resolution is needed. 

On the other hand atomic hydrogen has a line at 21 cm in an easily accessible, 
and well protected, region of the radio spectrum. 
However the observations are difficult because the intensity is weak 
and above all because there is a contamination by interstellar hydrogen  
along the same lines of sight. For a long time these difficulties have limited 
the \HI observations of AGB sources to $o$ Cet (Bowers \& Knapp 1988, BK1988). 
Nevertheless we recently readdressed this question with the upgraded 
Nan\c{c}ay Radiotelescope and could detect this line in emission towards 
several AGB sources. We have already presented our results on four 
of them: RS Cnc (G\'erard \& Le~Bertre 2003, Paper I), 
EP Aqr and Y CVn (Le~Bertre \& G\'erard 2004, Paper II) and 
X Her (Gardan et al. 2006, Paper III). We detected these four AGB
sources with a good signal-to-noise ratio 
and showed that the line profiles are composite. The individual 
components have Gaussian-like profiles, although in some cases a rectangular 
profile could be fitted as well. The \HI emissions are spread over 
a velocity range confined to the total CO velocity extent. However, we noted 
that sometimes the central velocities in CO and \HI do not coincide exactly. 
Furthermore, in Paper I, we argued that atomic 
hydrogen is present already in the stellar atmosphere of the central star 
(RS Cnc) which might even be devoid of molecular hydrogen. In Paper II 
we could resolve angularly two circumstellar 
shells, and show that \HI may be present at large distances from the central 
stars, 0.5 pc or more for EP Aqr. 
The \HI data display a large scale asymmetry suggesting that asymmetric  
structures may show up early in the AGB phase (EP Aqr \& X Her), 
well before the planetary nebula phase.
In the case of the source associated to Y CVn, the \HI data even seem  
to probe a large region extending out to where the stellar outflow 
is interacting with the Interstellar Medium (ISM). 

It thus appears that the \HI emission line profiles, when observed with 
a good signal-to-noise ratio and a fair removal of the ISM contamination, 
may reveal important clues on the physics of stellar outflows and 
their kinematics. Also, as atomic hydrogen has a long lifetime in the ISM, 
it can be used to trace circumstellar envelopes out to large radial distances. 
In the present paper, 
we give results on 22 new sources for which we have now good quality spectra, 
or good upper limits. The main emphasis of our programme is put on AGB sources
and we have tried to select a variety of representative specimens in terms of 
variability types, chemical composition, mass loss rates, etc. 
In addition, we have also included two planetary nebulae 
in order to sample more completely the post-main-sequence 
evolution of stars of low and intermediate masses. Two of our sources 
have already been observed in \HI with the Very Large Array (VLA), 
$o$ Ceti (BK1988) and NGC\,7293 (Rodr{\'\i}guez et al. 2002, R2002). 
This allows a useful comparison between single-dish and interferometric data.

\section{OBSERVATIONS}

%% In a manner similar to \objectname authors can provide links to dataset
%% hosted at participating data centers via the \dataset{} command.  The
%% second curly bracket argument is printed in the text while the first
%% parentheses argument serves as the valid data set identifier.  Large
%% lists of data set are best provided in a table (see Table 3 for an example).
%% Valid data set identifiers should be obtained from the data center that
%% is currently hosting the data.

\subsection{Observing Procedure}\label{obs}

The observations have been performed with the Nan\c cay Radiotelescope (NRT). 
The NRT is a meridian instrument with a rectangular effective aperture of 
160\,m$\,\times$\,30\,m. The HPBW at 21 cm is 4$'$ in the East-West direction 
and 22$'$ in the North-South direction. It has been upgraded 
(van Driel et al. 1996) and reopened to observations in 2001. 
The point source efficiency 
is 1.4 K\,Jy$^{-1}$ at 21 cm and the total system noise $\sim$ 35\,K. 
The autocorrelator has 8192 channels split in 4 banks of 2048, recording 
the linear (PA = 0 and 90$^{\circ}$) and circular polarisations.
For our programme we consider only the total intensity (Stokes parameter I). 

Two modes of observation have been used. In the frequency-switch (f-switch) 
mode,  the data are acquired on source with a spectral resolution of 
0.08 \kms ~and a bandwidth of 160 \kms. This mode is very useful to check 
the level of interstellar \HI emission in the direction of the source. 
A first indication of this level can be obtained by consulting the existing 
galactic \HI surveys (e.g. Hartmann \& Burton 1997), but their sensitivity 
limits are too high to readily select directions in which it might be possible 
to detect circumstellar \HI with a priori minimum background confusion. 
In the course of our programme, we characterize qualitatively the confusion
by ``weak'', ``mean'' and ``high''. These indications are reported 
in Table~\ref{HItab_a} (col. 2, see Sect.~\ref{Sample}). 
They are based on Hartmann \& Burton's results that 
are obtained with a single dish of HPBW $\sim 0.5^{\circ}$, as well as on our 
own f-switch observations. It should be noted that the impact of confusion 
depends on the telescope beam size and on the observing strategy, 
in particular single-dish versus interferometric mode. 
Exceptionnally \HI from the source can be seen directly on the f-switch 
spectra (e.g. Y CVn, Paper II). 

For the circumstellar \HI observations proper we use the position-switch 
mode, in which the telescope field of view (f.o.v.) is alternated between 
the on-source position and 2 off-source positions, placed symmetrically 
in the East-West direction. 
We refer to Paper II for a more detailed description of our observing 
procedure. It has proven to be most effective 
in separating the circumstellar emission from the interstellar one. 
A reason for this success comes from the fan beam shape ($4'\times22'$) of 
the telescope that allows a good East-West discrimination of the 
circumstellar envelope against the galactic \HI background. 
It should be noted that this procedure works perfectly when the intensity of 
the background 
varies linearly in right ascension across the region centered on the source. 
However, if the background presents a quadratic variation, or variations 
of higher {\it even} orders, spurious spectral features may appear. 
Such features generally grow rapidly with increasing 
separation between the 2 off-positions, which allows to identify them. 
 
In general we selected a bandwidth of 80 \kms ~with a spectral resolution 
of 0.04 \kms. For some sources where a large expansion velocity is expected 
from CO line data or where galactic confusion affected part of the spectrum, 
we have used a bandwidth of 160 \kms. 
The data are then smoothed to give a final spectral resolution of 0.32 \kms.
Except when mentioned otherwise, this resolution is used by default in the 
figures (Sect.~\ref{Sample}).

Finally, the source sizes are often comparable or larger than the NRT beam 
(4$'$ in the East-West direction). The intensity increases with beam throw 
and then reaches a maximum for $\pm$ n beams when both off-positions 
(East and West) lie entirely outside the source. 
We then reconstruct the total intensity ($s_{tot}$) by summing up 
the source fluxes in the different off-positions ($s_{k}$) and 
in the central position ($s_0$):\\
\begin{eqnarray}
s_{tot}=\sum_{k=-n+1}^{n-1}s_k
\end{eqnarray}
This is done by combining the position-switch spectra obtained at 
$\pm$ k beams ($C_{k}EW$) in the following expression:
\begin{eqnarray}
s_{tot}=(2n-1){\times}C_{n}EW - 2\sum_{k=1}^{n-1}C_{k}EW
\end{eqnarray}
n being such that the intensity in 2 successive position-switch 
spectra is the same, i.e. $C_{n+1}EW = C_{n}EW$. 
The radius of the source can then be estimated to $(n-1)\times4'\pm2'$.
Equ.~(2) assumes 
that the background varies linearly across the sky. On the other hand 
it makes no assumption on the brightness distribution of the source. 
For an unresolved source ($\phi \le 4'$), Equ. (2) reduces to:\\
\begin{eqnarray}
s_{tot}=C_{1}EW=C_{2}EW=C_{3}EW, etc. 
\end{eqnarray}
In the figures (next Section), the position-switch spectra ($C_{1}EW$, 
$C_{2}EW$, etc.) are shown in the upper panels, and their averages 
and the final spectra ($s_{tot}$) are displayed at the bottom 
(continuous and dashed curves, respectively). 

The average spectrum has a physical meaning only for an unresolved source 
(Equ. 3). However, the resulting signal-to-noise ratio is in general higher 
than for the total spectrum, so that spectral details are sometimes better 
seen on the average profile (see, e.g. below, Y UMa or R Peg). The difference 
between the two sets of spectra is the result of the source resolution 
by the 4$'$ beam.

\subsection{Sample of Sources}\label{Sample}

Table \ref{Sample_tab} lists the sources discussed in the present 
paper. We give the common name used in this work (col. 1), the 
IRAS identifier (col. 2), and some basic data. Except if mentioned otherwise, 
the spectral type (col. 3), the variability type (col. 4) and the period 
(col. 5) are taken from SIMBAD, directly or through the VizieR Service. 
The distances (col. 6) are in general deduced from the Hipparcos 
parallaxes. Next (col. 7), we give an estimate of the stellar effective 
temperature, \Teff. This datum is important because hydrogen in the 
stellar atmosphere and in the inner circumstellar envelope is expected to 
be mostly atomic for \Teff $>$ 2\,500 K and molecular for \Teff $<$ 2\,500 K
(Glassgold \& Huggins 1993, GH1983). Finally we give the LSR radial velocity
(\Vlsr, col. 8), the expansion velocity (\Vexp, col. 9), and the mass loss 
rate, estimated in general from CO rotational line data. These data 
are useful because both the central velocity and the velocity extent 
of the \HI lines are expected to be close to those given by the CO lines. In 
general CO data from different observers are consistent within $\pm$\,2 \kms. 
When no CO data are available ($\delta^2$ Lyr, $\rho$ Per, $\alpha^1$ Her, 
U CMi, NGC 6369), we adopt equivalent parameters from optical observations. 
Unfortunately, optical data seem less reliable than CO data ($\pm$\,5 \kms), 
perhaps in part due to the stellar variability. Except for U CMi, 
circumstellar envelopes have been detected around all sources of our sample, 
either in the CO radio lines, or in optical lines.

{
\begin{deluxetable}{lccccccccc}
\tabletypesize{\scriptsize}
\tablecaption{Observed sources in right ascension order 
and basic data.\label{Sample_tab}}
% Data are from the literature.
\tablewidth{0pt}
\tablehead{
\colhead{Source} & \colhead{IRAS name} & \colhead{Spectral} & \colhead{Variability} & 
\colhead{Period} & \colhead{d} & \colhead{\Teff} & \colhead{\Vlsr} & \colhead{\Vexp} & 
\colhead{\.M}\\
\colhead{ } & \colhead{ } & \colhead{type} & \colhead{type} & \colhead{(days)} & 
\colhead{(pc)} & \colhead{(K)} & \colhead{(\kms)} & \colhead{(\kms)} & \colhead{(\Msold)}
}
\startdata
%\hline 
WX Psc         & 01037+1219   & M8-M10:    & Mira   & 660    &  650 & $<$\,2500    &    +9 &    23    & 2   10$^{-5}$\\
$o$ Cet        & 02168$-$0312 & M7IIIe+Bep & Mira   & 332    &  128 & $\sim$\,2700 &   +46 &     7    & 2.5 10$^{-7}$\\
$\rho$ Per     & 03019+3838   & M4II       & SRb    &  40    &  100 & 3576         &   +25 & $\leq$8.5& 1   10$^{-8}$\\
NML Tau        & 03507+1115   & M6e-M10e   & Mira   & 462    &  245 & 2000-3000    &   +35 &    19    & 3   10$^{-6}$\\
S CMi          & 07299+0825   & M6e-M8e    & Mira   & 332    &  365 & 2500-3100    &   +52 &     3    & 2   10$^{-7}$\\
U CMi          & 07386+0829   & M4e        & Mira   & 414    &  191 & 3400         &   +42 &    ...   & ... \\
RV Hya         & 08372$-$0924 & M5II       & SRc    & 116    &  318 & 3420         & $-$43 &     5    & 3   10$^{-7}$\\
U Hya          & 10350$-$1307 & C5II       & SRb    & 450    &  162 & 2965         & $-$31 &     7    & 9   10$^{-8}$\\
Y UMa          & 12380+5607   & M7II-III   & SRb    & 168    &  313 & 3100         &   +19 &     5    & 2   10$^{-7}$\\
RY Dra         & 12544+6615   & C4,5J      & SRb    & 200    &  488 & 2810         &  $-$4 &    13    & 1.5 10$^{-7}$\\
RT Vir         & 13001+0527   & M8III      & SRb    & 155    &  138 & 3034         &   +17 &     9    & 1   10$^{-7}$\\ 
W Hya          & 13462$-$2807 & M7.5e-M9ep & SRa    & 361    &  115 & 2500         &   +41 &     8    & 2   10$^{-7}$\\
$\alpha^1$ Her & 17123+1426   & M5Ib-II    & SRc    & ...    &  117 & 3285         & $-$17 &    10    & 1   10$^{-7}$\\ 
NGC 6369       & 17262$-$2343 & PN         & ...    & ...    & 1550 & $>$\,10000   & $-$90 &    41    & ... \\
$\delta^2$ Lyr & 18527+3650   & M4II       & ?      & 60-120 &  275 & 3460         &  $-$7 & $\leq$10 & 5   10$^{-8}$\\
Z Cyg          & 20000+4954   & M5e-M9e    & Mira   & 264    &  490 & $<$\,3300    & $-$148&     4    & 4   10$^{-8}$\\
NGC 7293       & 22267$-$2102 & PN         & ...    & ...    &  200 & $>$\,10000   & $-$23 &    32    & ... \\
R Peg          & 23041+1016   & M6e-M9e    & Mira   & 378    &  350 & 2300-2900    &   +24 &     9    & 5   10$^{-7}$\\
AFGL 3068      & 23166+1655   & C          & Mira   & 696    & 1140 & 1800-2000    & $-$31 &    14    & 1   10$^{-4}$\\
AFGL 3099      & 23257+1038   & C          & Mira   & 484    & 1500 & 1800-2000    &   +47 &    10    & 8   10$^{-6}$\\
TX Psc         & 23438+0312   & C5II       & Lb     & ...    &  233 & 3115         &   +13 &    10    & 2   10$^{-7}$\\
R Cas          & 23558+5106   & M7IIIe     & Mira   & 430    &  107 & $<$\,3000    &   +25 &    12    & 5   10$^{-7}$\\
\enddata
\end{deluxetable}
}

In Table \ref{HItab_a}, we give the parameters characterizing 
the \HI emissions. In general these emissions can be fairly well fitted 
with a Gaussian. We derive the LSR central velocity, 
V$_{\rm cent.}$ (col. 4), a full width at half maximum, FWHM (col. 5), 
and a peak intensity, F$_{\rm peak}$ (col. 6), obtained from an unconstrained 
positive Gaussian fit to the integrated flux ($s_{tot}$; Equ. 2). 
In the cases of composite profile (Y UMa, R Peg), two such Gaussians are 
also fitted to the average profile in the lower panel of the 
corresponding figure. When part of the spectrum is affected 
by interstellar confusion it has been excluded from the range used for the 
Gaussian fit. In such cases, the part of the spectrum which is affected is 
always that closest to 0 \kms ~LSR. The Flux (col. 7) is estimated from 
the integral of the $s_{tot}$ Gaussian fit and therefore only takes 
into account the extension of the source in right ascension. 
Finally the hydrogen mass, 
\MHI ~(col. 8), is derived from the Flux and for the distance, d 
(Table \ref{Sample_tab}). We assume negligible absorption of the circumstellar 
\HI emission by foreground galactic gas and self-absorption, 
and thus may slightly underestimate the \HI mass.

\subsubsection{AGB Sources}\label{Sample_AGB}

Stars on the AGB present a large variety of properties in terms of effective 
temperature, luminosity, chemistry, variability, etc., probably reflecting 
sources of different ages, initial masses, etc. The observed mass loss rate 
ranges from 10$^{-8}$ to 10$^{-4}$ \Msold. In fact AGB stars seem to undergo 
mass loss at a rate which may vary on timescales ranging from a few years 
to perhaps 10$^{6}$ years. We have tried to select oxygen-rich and carbon-rich 
sources with different variability types and different mass loss rate 
estimates. Our sample covers also a wide range of stellar temperatures.

%% Observe the use of the LaTeX \label
%% command after the \subsection to give a symbolic KEY to the
%% subsection for cross-referencing in a \ref command.
%% You can use LaTeX's \ref and \label commands to keep track of
%% cross-references to sections, equations, tables, and figures.
%% That way, if you change the order of any elements, LaTeX will
%% automatically renumber them.

%% This section also includes several of the displayed math environments
%% mentioned in the Author Guide.

%% The \notetoeditor{TEXT} command allows the author to communicate
%% information to the copy editor.  This information will appear as a
%% footnote on the printed copy for the manuscript style file.  Nothing will
%% appear on the printed copy if the preprint or
%% preprint2 style files are used.

\subsubsubsection{Oxygen-rich AGB Sources}

We discuss first $\delta^2$ Lyr, whose variability type is not defined, 
and then the 6 oxygen-rich Semi-Regular variables of our sample. 

{\bf $\delta^2$ Lyr} (HR 7139) is a relatively warm AGB star (M4II), which 
is barely variable ($\Delta$~V~$\sim$ 0.2 mag., Bakos \& Tremko 1991).
We adopt the effective temperature determined by Sudol et al. (2002) from 
near-infrared (J and H) interferometry, 3\,460 $\pm$ 160~K. 
From the profiles of photospheric lines affected by circumstellar 
absorption, Sanner (1976) finds evidence of an outflow. This source 
has not been detected in any radio line, but it has an IRAS extended 
counterpart (Young et al. 1993a).  On the other hand, the ISO (3-40 $\mu$m) 
spectrum (Heras et al. 2002) shows no evidence of dust. In Table 1, we adopt 
the parameters of its wind from Sanner (1976). 

\begin{figure} 
\includegraphics[angle=0,scale=.42]{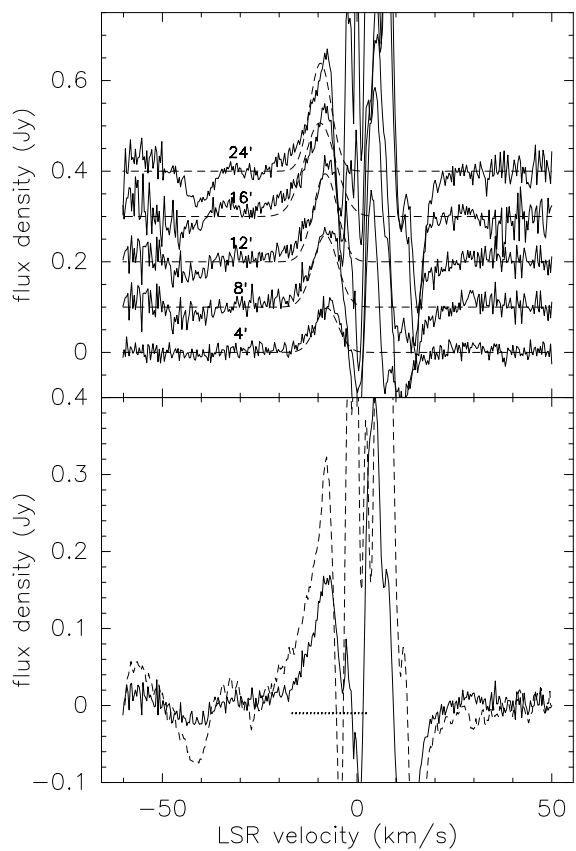}
\caption[]{Top panel: $\delta^2$ Lyr spectra obtained in the position-switch 
mode.
%with off-positions at $\pm 4'$, $\pm 8'$, $\pm 12'$, $\pm 16'$, and $\pm 24'$.
The off-positions, in this figure and ensuing ones, are indicated in 
bold characters above each spectrum. For clarity, the individual spectra 
%in this figure and the following ones 
have been displayed with vertical offsets of 0.05 or 0.1 Jy.
Bottom panel: average of these 5 spectra and space integrated intensity 
scaled by a factor 1/2. The horizontal dashed bar indicates 
the velocity range expected for circumstellar matter (Sanner 1976).}
\label{delta2Lyr_HI_RA_av}
\end{figure}

%We have obtained 21 hours of integration with off-positions at $\pm 4'$, 
%$\pm 8'$, $\pm 12'$, $\pm 16'$, and $\pm 24'$ 
The \HI spectra are presented in 
Fig.~\ref{delta2Lyr_HI_RA_av}.
In this figure, and in the following ones, the horizontal bar 
represents the velocity extent (\Vlsr $\pm$ \Vexp) 
expected from the wind parameters given in Table \ref{Sample_tab} 
(full line in the case of a CO radio-line determination, 
dotted line in the case of an optical determination).
The contamination is serious especially at V $>$ $-$5 \kms. However, each 
spectrum clearly shows an \HI emission that is increasing with offset up to 
$\pm$ 16$'$. This indicates that the source has a diameter of $\sim$ 24$'$. 
The central velocity of the profile and the FWHM agree with those obtained by 
Sanner (1976).
The intensity has been integrated over the 7 NRT beams that contribute to 
the signal. At a distance of 275 pc, it translates to $\sim$ 0.075 \Msol 
~in atomic hydrogen. As the source has a size of $\sim$ 24$'$ in the East-West 
direction and as the NRT beam is 22 $'$ in the North-South direction, we may 
have missed part of the \HI emission (see Sect.~\ref{spatialdistributions} 
for a corrected \HI mass).

\clearpage

\begin{figure} 
\includegraphics[angle=0,scale=.42]{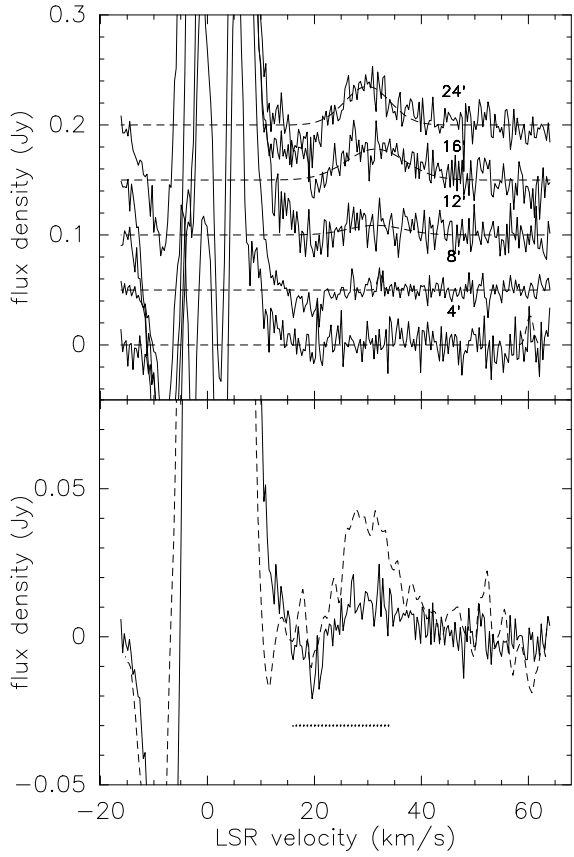}
\caption[]{Top panel: $\rho$ Per spectra obtained in the position-switch 
mode.
% with off-positions at $\pm 4'$, $\pm 8'$, $\pm 12'$, $\pm 16'$ and $\pm 24'$.
Bottom panel: average of the 5 spectra and space integrated intensity scaled 
by a factor 1/5. The horizontal dashed bar indicates the velocity range 
expected for circumstellar matter (Sanner 1976).}
\label{rhoPer_HI_RA_av}
\end{figure}

\clearpage

{\bf $\rho$ Per} (HR 921) has not been detected in CO, nor in any other 
radio line. Sanner (1976) finds evidence of an outflow with parameters 
(\Vlsr, \Vexp, \.M) given in Table 1. Mauron \& Guilain (1995) detected 
circumstellar Na\,{\sc {i}} at 5$''$ from the central star (i.e. $\sim$ 
500 AU), confirming that the star is losing matter. However, 
the infrared spectrum obtained by ISO shows no evidence of dust 
around $\rho$ Per (Heras et al. 2002). 
We adopt the effective temperature from Dumm \& Schild (1998).

%We have obtained 50 hours of integration with off-positions at $\pm 4'$, 
%$\pm 8'$, $\pm 12'$, $\pm 16'$ and $\pm 24'$ (Fig.~\ref{rhoPer_HI_RA_av}).  
An \HI emission is barely detected in the spectrum obtained with off-positions 
at $\pm 4'$, whereas it is clearly detected at $\pm 16'$, and at 
$\pm 24'$ with basically the same intensity (Fig.~\ref{rhoPer_HI_RA_av}). 
This indicates that the \HI emission is resolved and has a size corresponding 
to about 7 NRT beams in the East-West direction, i.e. $\sim$ 24$'$. 
Galactic confusion clearly affects the \HI profile near 20 \kms 
~at $\pm$ 16 and 24$'$. 
The blue wing of the \HI profile is cut down, which probably 
results in a displacement of the \HI velocity centroid.
%in an apparent redshift of the centroid velocity.

\clearpage

\begin{figure} 
\includegraphics[angle=0,scale=.42]{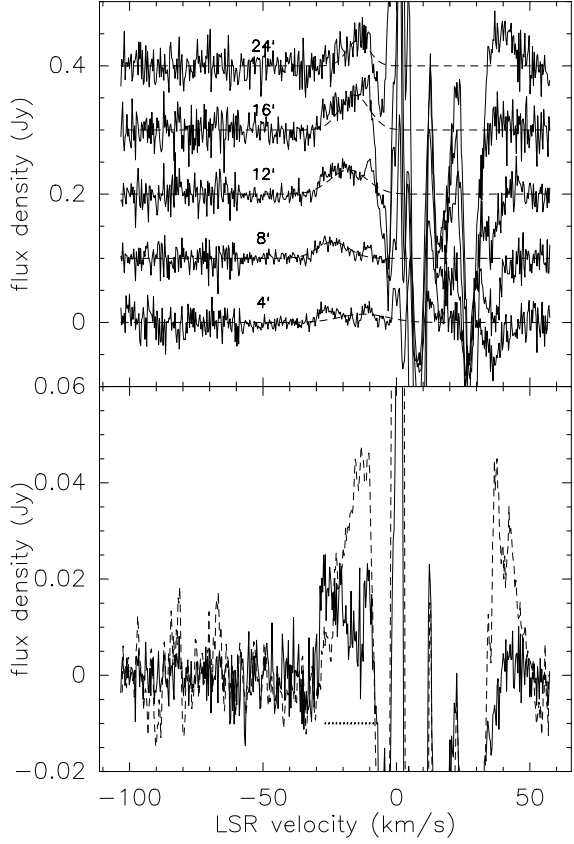}
\caption[]{Top panel: $\alpha^1$ Her spectra obtained in the position-switch 
mode.
%with off-positions at $\pm 4'$, $\pm 8'$, $\pm 12'$, $\pm 16'$, and $\pm 24'$.
Bottom panel: average of these 5 spectra and space integrated 
intensity scaled by a factor 1/5. The horizontal dashed bar indicates 
the velocity range expected for circumstellar matter (Deutsch 1956).}
\label{alphaHer_HI_RA_av}
\end{figure}

\clearpage

{\bf $\alpha^1$ Her} (HR 6406)
%However, 
is known to undergo mass loss since Deutsch (1956) showed that 
circumstellar lines are present in absorption in the spectrum of its 
companion, at an angular distance of 5$''$. We adopt his wind parameters. 
%Reimers (1977, 1978) finds also evidence of mass loss. 
Like $\rho$ Per, it was not detected in any 
radio line (e.g. Heske 1990). 
Circumstellar K\,{\sc {i}} and 
Na\,{\sc {i}} were also detected in emission by Mauron \& Caux (1992).
We adopt the effective temperature determined by Perrin et al. (2004a) from 
2 $\mu$m interferometric data (\Teff\,= 3\,285 $\pm$ 89 K). 

%We have obtained 46 hours of integration with off-positions at $\pm 4'$, 
%$\pm 8'$, $\pm 12'$, $\pm 16'$, and $\pm 24'$ (Fig.~\ref{alphaHer_HI_RA_av}). 
The \HI emission is increasing with offset (Fig.~\ref{alphaHer_HI_RA_av}), 
which indicates that the source 
has a diameter of the order of 16$'$. The central velocity of the profile and 
its width agree with the parameters obtained by Deutsch (1956). The 
interstellar confusion is high for V $> -$5 \kms, and seems even to affect 
the red wing of the profile obtained at $\pm 16'$. An emission component 
stands out at $-$10 \kms. As it can be seen on the individual position-switch 
spectra at  $\pm 4'$, $\pm 8'$, and $\pm 12'$ it is probably real.

{\bf RV Hya} has been detected in CO by Winters et al. (2003). 
We adopt their parameters for the wind characteristics. With the temperature 
scale of normal giants (Ridgway et al. 1980) and an M5 spectral type, 
the effective temperature should be around 3\,400\,K.

\begin{figure} 
\includegraphics[angle=270,scale=.28]{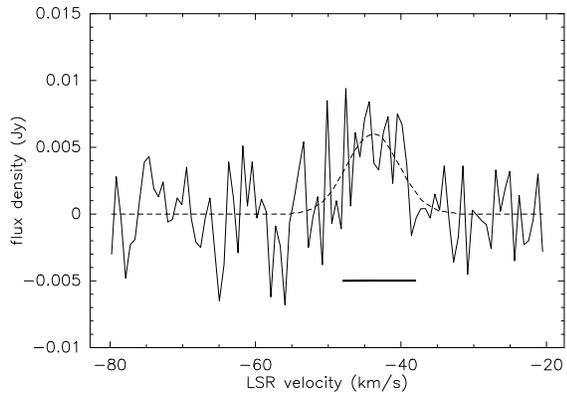}
\caption[]{Average spectrum of RV Hya obtained in the 
position-switch mode. The spectral resolution is 0.64 \kms.}
\label{RVHya_HI_av}
\end{figure}

%We have obtained 35 hours of integration with off-positions at $\pm 4'$, 
%$\pm 8'$, and $\pm 12'$. 
The \HI line is marginally detected on the average 
(Fig.~\ref{RVHya_HI_av}). It does not seem to be angularly resolved. 

\clearpage

\clearpage

{\bf Y UMa} is associated to an IRAS extended source (Young et al. 1993a). 
It has been observed in CO (2--1) by Knapp et al. (1998) and their wind 
characteristics are given in Table 1. These estimates agree with those 
obtained by Olofsson et al. (2002). From the temperature scale of normal 
giants of type later than M6 (Perrin et al. 1998), 
we adopt an effective temperature of about 3\,100\,K. 

\begin{figure} 
\includegraphics[angle=0,scale=.38]{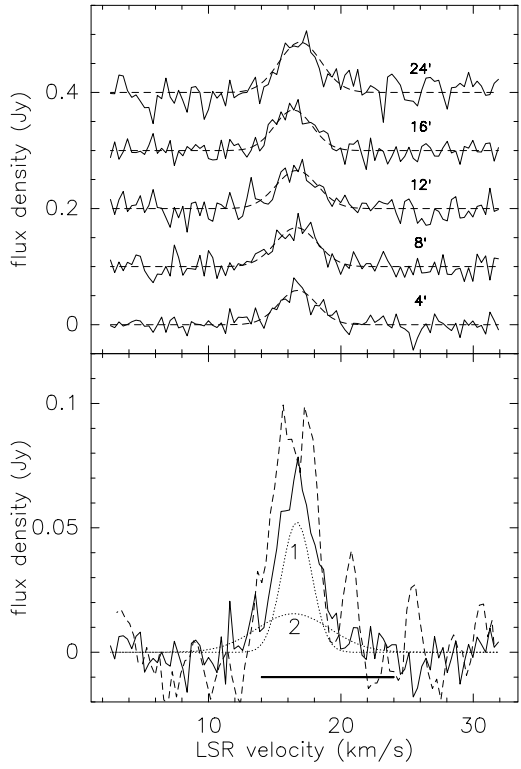}
\caption[]{Top panel: Y UMa spectra obtained in the position-switch 
mode.
%with off-positions at $\pm 4'$, $\pm 8'$, $\pm 12'$, $\pm 16'$, and $\pm 24'$.
Bottom panel: average of these 5 spectra 
and space integrated intensity. Gaussian fits to the average 
are shown to illustrate the decomposition of the profile in 2 components.}
\label{YUMa_HI_RA_av}
\end{figure}

%We have obtained 28 hours of integration with off-positions at $\pm 4'$, 
%$\pm 8'$, $\pm 12'$, $\pm 16'$, and $\pm 24'$ (Fig.~\ref{YUMa_HI_RA_av}). 
The \HI emission is spatially extended ($\phi \sim$ 8$'$, 
Fig.~\ref{YUMa_HI_RA_av}).  
We note that the central velocity of the \HI feature is shifted by $\sim$ 
2 \kms ~with respect to the CO emission and is narrower (Knapp et al. 1998). 
The profile is probably composite with a narrow component overimposed 
on a broad faint one. A detailed examination of the separate 
position-switch data shows that the \HI emission 
is asymmetric and slightly shifted West.

\clearpage

\clearpage

{\bf RT Vir} is also associated to an IRAS extended source (Young et al. 
1993a). We adopt the wind characteristics from Knapp et al. (1998). However, 
Olofsson et al. (2002) derive a larger mass loss rate estimate ($\sim$ 
5$\times$10$^{-7}$ \Msold, at 170 pc). It has been detected in the thermal 
(v=0, J=2-1) SiO line by Gonz\'alez Delgado et al. (2003) at a slightly 
different velocity (\Vlsr $\sim$ 18.6 \kms). This source was also detected in 
the OH maser lines at 1\,612, 1\,665 and 1\,667 MHz by Etoka et al. (2003).
The effective temperature is taken from Dumm \& Schild (1998).

\begin{figure} 
\includegraphics[angle=0,scale=.42]{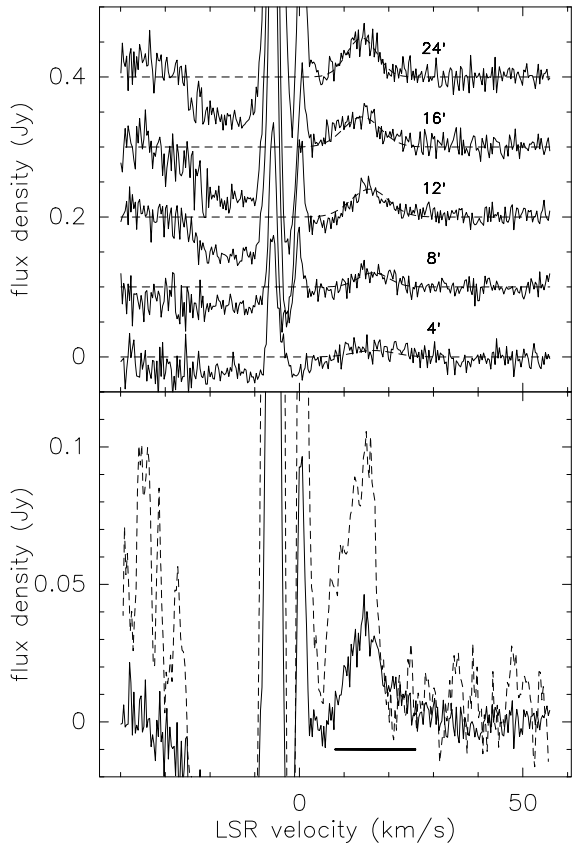}
\caption[]{Top panel: RT Vir spectra obtained in the position-switch mode.
%with off-positions at $\pm 4'$, $\pm 8'$, $\pm 12'$, $\pm 16'$, and $\pm 24'$.
Bottom panel: average of these 5 spectra 
and space integrated intensity scaled by a factor 1/2.}
\label{RTVir_HI_RA_av}
\end{figure}

%We have obtained 24 hours of integration (Fig.~\ref{RTVir_HI_RA_av}) with 
%off-positions at $\pm 4'$, $\pm 8'$, $\pm 12'$, $\pm 16'$, and $\pm 24'$. 
The \HI source is extended, $\phi \sim 24'$ (Fig.~\ref{RTVir_HI_RA_av}).
A detailed examination of the position-switch data shows that the \HI 
emission is asymmetric and shifted East by $\sim 4'$.

\clearpage

{\bf W Hya} is sometimes classified as a Mira (e.g. Young 1995). 
It is associated to an IRAS extended source (Young et al. 1993a). 
We adopt the wind characteristics from Knapp et al. (1998).
These parameters agree with those obtained by Young (1995) and Olofsson et al. 
(2002). The H$_2$O line at 557 GHz was detected by the Odin satellite 
and gives consistent parameters for the wind (Justtanont et al. 2005).
The circumstellar shell was also detected in K\,{\sc {i}}
fluorescence by Guilain \& Mauron (1996). Like RT Vir, it has been detected 
in the OH maser lines at 1612, 1665 and 1667 MHz (Etoka et al. 2003). 
An effective temperature of 2\,500 $\pm$\,190\,K was obtained by Haniff et al. 
(1995) from optical interferometry. The stellar disk was also resolved 
in the radio continuum by Reid \& Menten (1997) who derived a (radio) 
brightness temperature of 1\,500 $\pm$\,570\,K.

\begin{figure}
\includegraphics[angle=0,scale=.42]{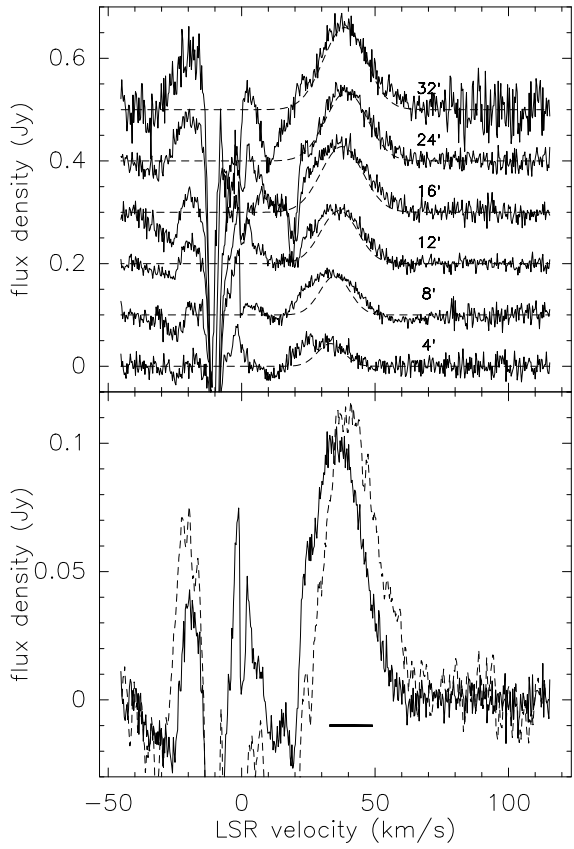}
\caption[]{Top panel: W Hya spectra obtained in the position-switch 
mode.
%with off-positions at $\pm 4'$, $\pm 8'$, $\pm 12'$, $\pm 16'$, 
%$\pm 24'$ and $\pm 32'$. 
Bottom panel: average of these 6 spectra
and space integrated intensity scaled by a factor 1/5.}
\label{WHya_HI_RA_av}
\end{figure}

%We have obtained 43 hours of integration with off-positions at $\pm 4'$, 
%$\pm 8'$, $\pm 12'$, $\pm 16'$, $\pm 24'$ and $\pm 32'$ 
%(Fig.~\ref{WHya_HI_RA_av}). 
The \HI source is clearly resolved with a maximum flux reached only at 
$\pm 16'$ so that we estimate its size to $\sim$ 24$'$ 
(Fig.~\ref{WHya_HI_RA_av}). The spectra obtained at $\pm16'$ and $\pm24'$ show 
a narrow dip at +20\,\kms ~(FWHM $\sim$ 3 \kms), revealing the presence 
of a small interstellar cloud at $\sim 20'$ East of W\,Hya.

\clearpage

We now discuss the 8 oxygen-rich Miras of our sample, in order of increasing 
spectral type, which corresponds approximately to increasing period and 
increasing observed mass loss rate. The spectral types of Miras are known 
to vary with phase, and accordingly the stellar effective temperatures, which 
makes them particularly uncertain.

{\bf U CMi} is probably one of the bluest Miras. As it has not been 
detected in any radio lines, we adopt the radial velocity determined 
optically (General Catalogue of Stellar Radial Velocities, GCRV). 
Adopting the effective temperature scale for Mira variables 
of Dyck et al. (1974), \Teff ~should be around 3400\,K.

\begin{figure} 
\includegraphics[angle=0,scale=.42]{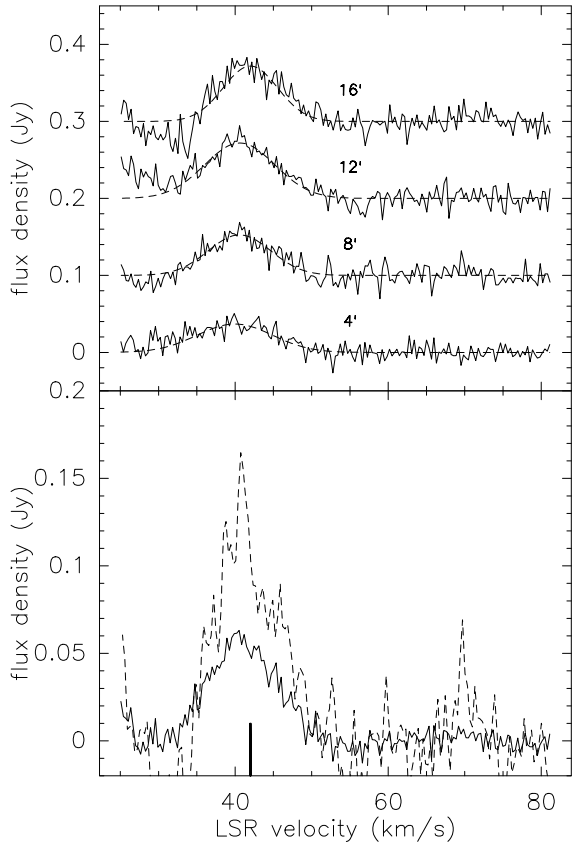}
\caption[]{Top panel: U CMi spectra obtained in the position-switch mode.
%with off-positions at $\pm 4'$, $\pm 8'$, $\pm 12'$, and $\pm 16'$. 
Bottom panel: average of these 4 spectra and space integrated intensity.}
\label{UCMi_HI_RA_av}
\end{figure}

%We have obtained 20 hours of integration with off-positions at $\pm 4'$, 
%$\pm 8'$, $\pm 12'$, and $\pm 16'$ (Fig.~\ref{UCMi_HI_RA_av}). 
As there is no estimate available for the expansion velocity, the stellar 
radial velocity is only marked with a vertical bar (Fig.~\ref{UCMi_HI_RA_av}). 
The \HI line is easily detected 
with a triangular profile centered at a velocity (40.9 \kms) very close 
to the stellar one (42 \kms). It is more intense for $\pm 8'$ 
than for $\pm 4'$ and we estimate its diameter to $\sim$ 8$'$. 
As for Y UMa, we observe that the \HI emission is not centered 
on the star position, and slightly shifted West.

\clearpage

{\bf Z Cyg} is remarkable for its high radial velocity suggesting it might 
belong to population II. It has been detected as an OH maser 
by Sivagnanam et al. (1989) and in the CO (3-2) line by Young (1995) 
from whom we adopt the outflow parameters and the distance. The effective 
temperature should be around 3\,300\,K or below (Dyck et al. 1974).

\begin{figure} 
\includegraphics[angle=270,scale=.28]{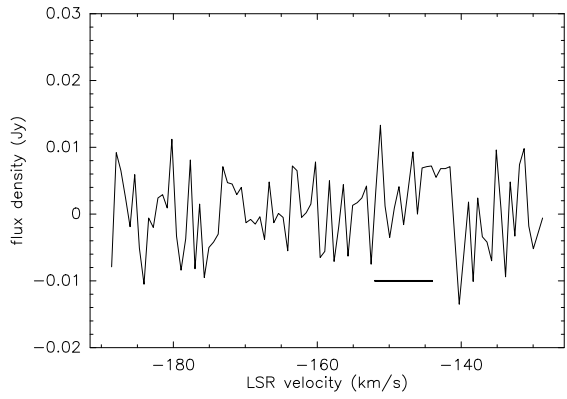}
\caption[]{Average spectrum of Z Cyg obtained in the 
position-switch mode. The spectral resolution is 0.64 \kms.}
\label{ZCyg_HI_av}
\end{figure}

%We have obtained 27 hours of integration with off-positions at $\pm 4'$, 
%$\pm 6'$, and $\pm 8'$ (Fig.~\ref{ZCyg_HI_av}). 
An \HI line is suspected 
at $\sim$ --146 \kms ~(Fig.~\ref{ZCyg_HI_av}) 
but we only set an upper limit to its flux.

\clearpage

{\bf S CMi} has been observed in CO (1$-$0) and (2$-$1) by Winters et al. 
(2003). We adopt their wind parameters and distance based on the 
period-luminosity relationship for Miras (Feast 1996). The effective 
temperature should stay most of the time above 2\,500 K (Dyck et al. 1974).

\begin{figure} 
\includegraphics[angle=0,scale=.42]{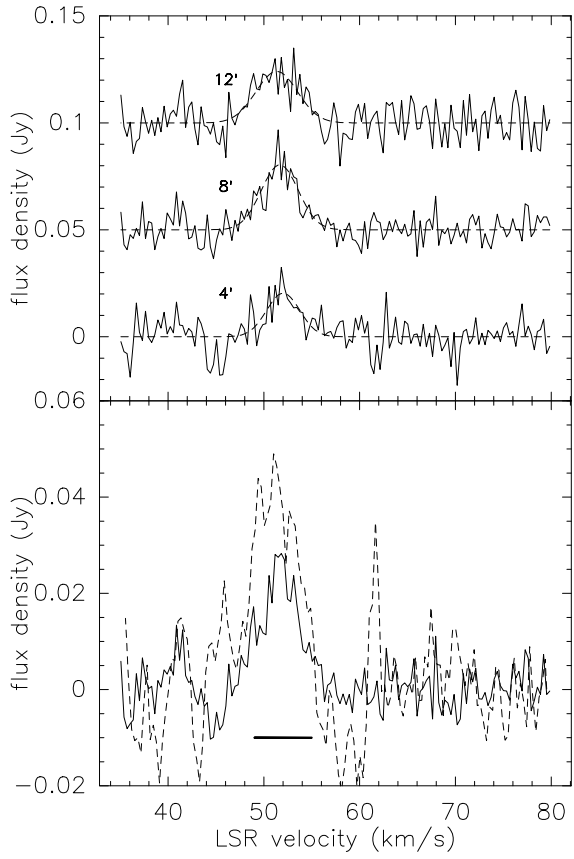}
\caption[]{Top panel: S CMi spectra obtained in the position-switch 
mode.
%with off-positions at $\pm 4'$, $\pm 8'$, and $\pm 12'$. 
Bottom panel: average of these 3 spectra and space integrated intensity.}
\label{SCMi_HI_RA_av}
\end{figure}

%We have obtained 27 hours of integration with off-positions at $\pm 4'$, 
%$\pm 8'$, and $\pm 12'$ (Fig.~\ref{SCMi_HI_RA_av}). 
The \HI line is easily detected (Fig.~\ref{SCMi_HI_RA_av})
and is slightly more intense for $\pm 8'$ than 
for $\pm 4'$. As there is no clear difference between $\pm 8'$ and $\pm 12'$
the diameter of the source should be $\sim$ 8$'$.

\clearpage

{\bf $o$ Cet} (Mira, HR 681) has been observed extensively. Its 
infrared spectrum shows a 10-$\mu$m silicate feature strongly in emission 
(e.g. Speck et al. 2000). The CO profiles 
are complex, possibly composite (Knapp et al. 1998; Winters et al. 2003).
Circumstellar K\,{\sc {i}} and Na\,{\sc {i}} were also 
detected by Mauron \& Caux (1992). CO (2$-$1) interferometric and 
K\,{\sc {i}} data show that the envelope is strongly 
asymmetric, perhaps bipolar, (Josselin et al. 2000).
We adopt a mass loss rate of 2.5$\times$10$^{-7}$ \Msold ~from the detailed 
modelling of CO rotational lines by Ryde \& Sch\"oier (2001). 
The central star has a hot companion, at approximately 0.5$''$ 
($\sim$ 65\,AU), surrounded by a disk accreting a part, $\approx$\,1\,\%, 
of the matter in the outflow (Reimers \& Cassatella 1985). 

On the basis of a large set of infrared interferometric data, Weiner (2004) 
finds that a 2\,200\,K H$_2$O shell surrounds a 2\,700\,K star (at a phase 
close to maximum). This estimate of the stellar temperature agrees with 
the Dyck et al. (1974) scale. On the other hand, Perrin et al. (2004b), 
using near-infrared interferometric data, derive an effective temperature 
in the range 3\,200-3\,600 K. These results illustrate that Mira atmospheres 
have complex structures, as predicted by 
%recent 
hydrodynamical models (e.g. Woitke et al. 1999). 
A prediction of the atomic hydrogen abundance 
in the stellar atmosphere and inner outflow would certainly require 
a detailed hydrodynamical and chemical modeling. 

$o$ Cet was already detected in \HI with the VLA (BK1988). These early 
observations suggest that approximately one third of the hydrogen is atomic 
when it leaves the star. The source is spatially resolved (radius $\sim 160''$)
and slightly offset to the North-West. However this large-scale asymmetry 
does not seem to be related to the bipolar flow ($\phi \sim 20''$) observed 
in CO by Josselin et al. (2000).

\begin{figure} 
\includegraphics[angle=0,scale=.42]{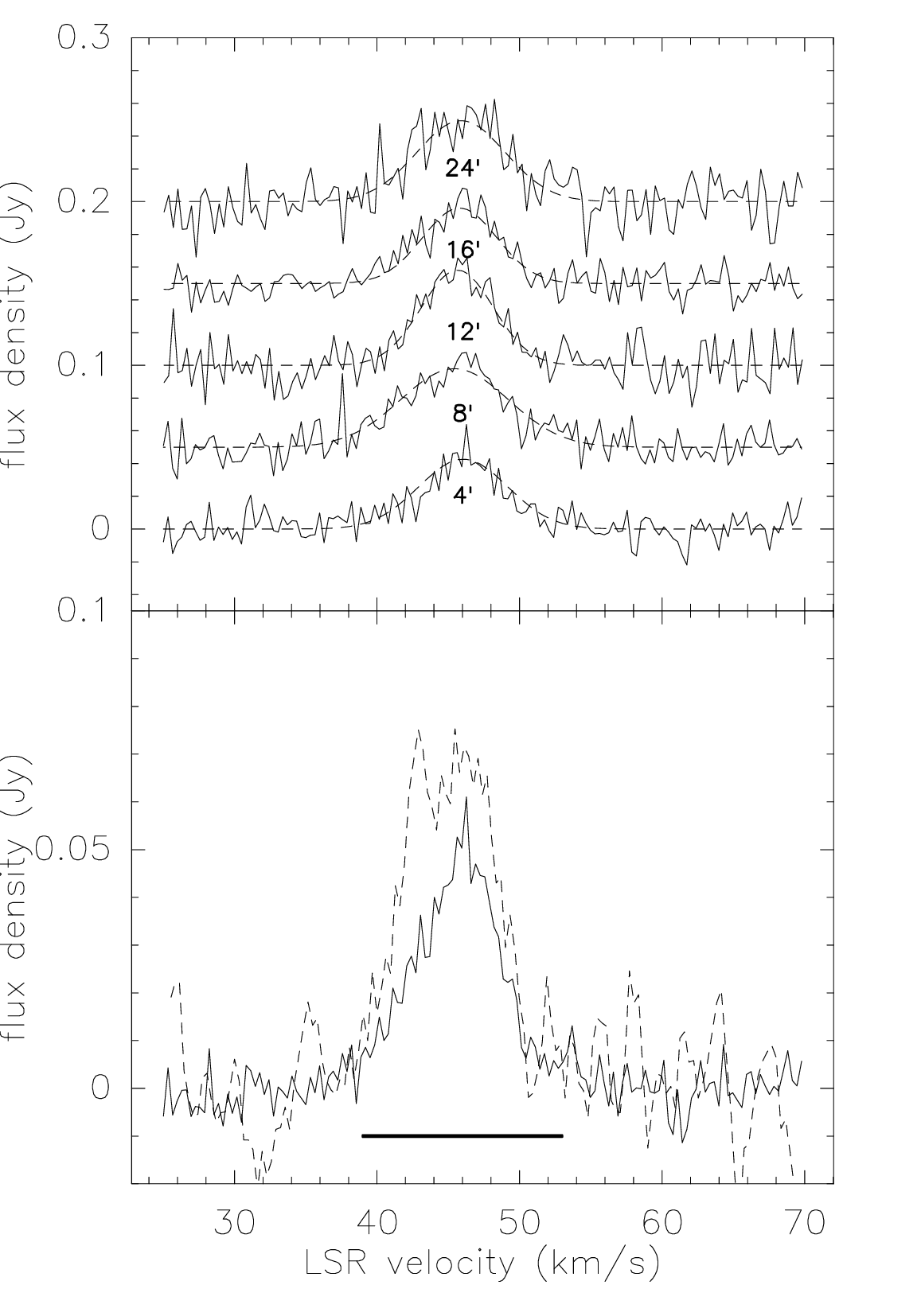}
\caption[]{Top panel: $o$ Cet spectra obtained in the position-switch mode. 
%with off-positions at $\pm 4'$, $\pm 8'$, $\pm 12'$, $\pm 16'$ and $\pm 24'$. 
Bottom panel: average of these 5 spectra and space integrated intensity.} 
\label{oCet_HI_RA_av}
\end{figure}

%We have obtained 34 hours of integration with off-positions at $\pm 4'$, 
%$\pm 8'$, $\pm 12'$, $\pm 16'$, and $\pm 24'$ (Fig.~\ref{oCet_HI_RA_av}). 
The signal that we obtained is slightly larger for $\pm 8'$ than for $\pm 4'$ 
and very stable for larger offsets (Fig.~\ref{oCet_HI_RA_av}). 
We estimate that the source is just 
resolved spatially with the NRT ($\phi_{\rm HI} \sim 8 \pm 4'$), consistent 
with the result of BK1988 and the extension ($\phi \sim 4.4'$) found by IRAS 
at 60 $\mu$m (Young et al. 1993a). The average line-profile is of high quality 
and resembles the triangular shape of the CO line-profiles 
(e.g. Knapp et al. 1998; Winters et al. 2003), with a small asymmetry 
shifting the peak (46.3 \kms) by $\sim$ 1 \kms ~to the red of 
the centroid velocity (45.7 \kms). Such a close 
agreement, in shape and in position, is exceptional in our sample.
But the most important fact maybe that this high-quality \HI line-profile 
clearly does not show the rectangular shape expected from an optically thin 
envelope with constant expansion velocity (Le Bertre \& G\'erard, Paper II). 
This point is further discussed in Sect.~{\ref{interpretation}}.

\clearpage

The integrated intensity is $\sim$ 0.56 Jy$\times${\kms} ~which (for a 
distance of 128 pc) translates into \MHI $\approx$ 2.2$\times$10$^{-3}$ \Msol. 
For comparison, BK1988 find 0.42 Jy$\times$\kms, 
which is in good agreement with our estimate.  

For a size of 8$'$ (0.30 pc) and an average expansion velocity of 
$\sim$4 \kms, the crossing time is $\sim$ 38$\times$10$^3$ years, and the 
average mass loss rate in \HI over this period is $\sim$ 0.6$\times$10$^{-7}$ 
\Msold. Assuming atomic hydrogen represents 3/4 of the total mass, we find 
(\.M)$_{\rm av}$ $\sim$ 0.8$\times$10$^{-7}$ \Msold ~which 
is one third of  
%agrees within a factor 3 with 
the present mass loss rate estimated from CO-lines modelling 
(Ryde \& Sch\"oier 2001). This result is in line with 
%supports 
the  interpretation of BK1988 that 1/3 of the hydrogen is leaving the star in
atomic form.

{\bf R Peg} has been observed in CO by Young (1995) and Winters et al. 
(2003). We adopt their wind characteristics. van Belle et al. (1996) 
obtained interferometric data at 2 $\mu$m and derived effective temperatures 
at phases close to maximum (2\,333 $\pm$ 100\,K), and close to minimum 
(2\,881 $\pm$ 153\,K).

\begin{figure} 
\includegraphics[angle=0,scale=.42]{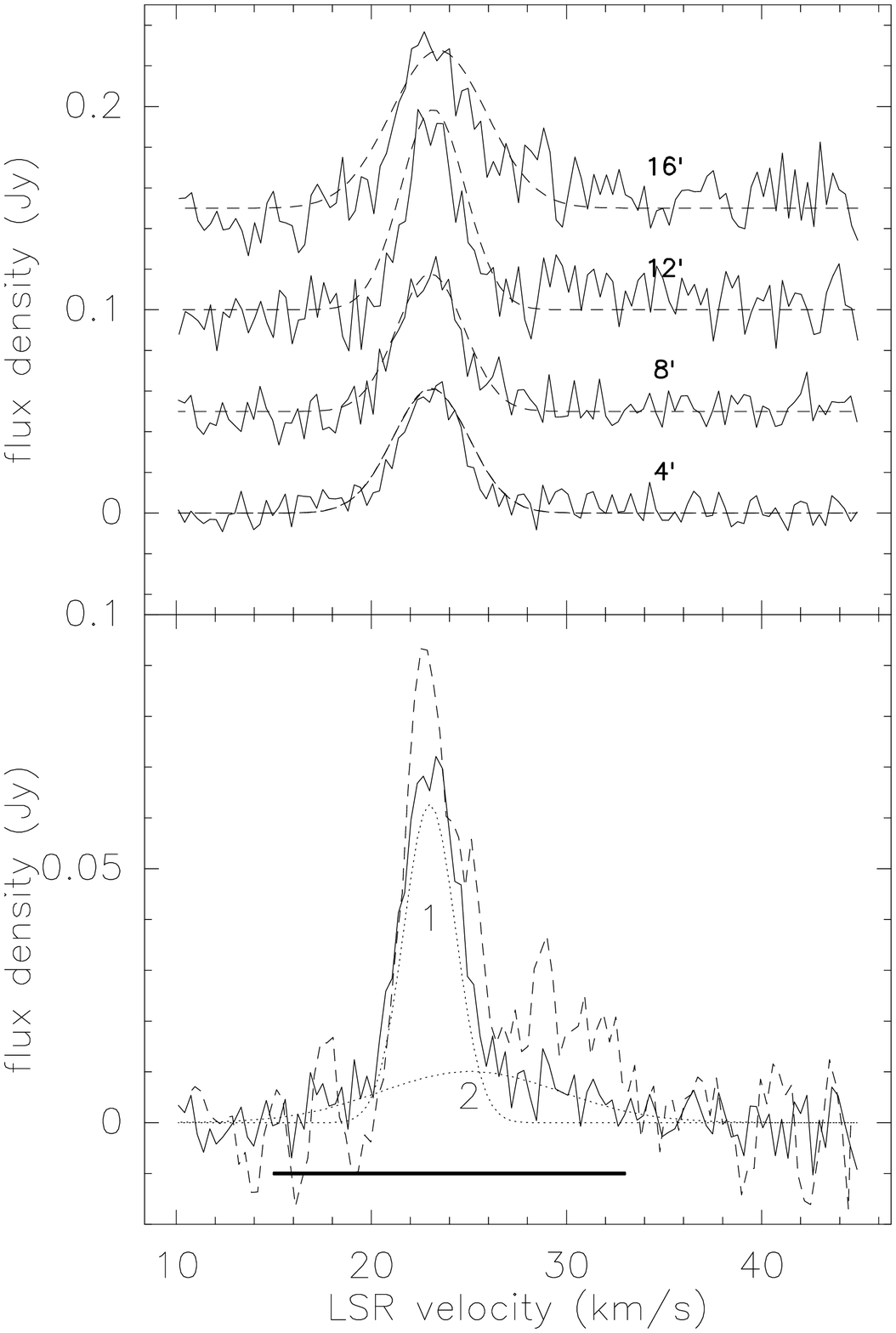}
\caption[]{Top panel: R Peg spectra obtained in the position-switch mode.
%with off-positions at $\pm 4'$, $\pm 8'$, $\pm 12'$, and $\pm 16'$. 
Bottom panel: average of these 4 spectra and space integrated 
intensity scaled by a factor 1/2. Gaussian fits to the average are 
shown to illustrate the decomposition of the feature in 2 components.}
\label{RPeg_HI_RA_av}
\end{figure}

%We have obtained 30 hours of integration with off-positions at $\pm 4'$, 
%$\pm 8'$, $\pm 12'$, and $\pm 16'$ (Fig.~\ref{RPeg_HI_RA_av}). 
As for Y UMa the \HI profile seems composite with a narrow component 
overimposed on a broader one (Fig.~\ref{RPeg_HI_RA_av}), and is 
reminiscent of the Y CVn \HI profile (Paper II). 
Furthermore, the narrow component (1) is spatially resolved 
with a size of $\sim$16$'$. However, the difference in 
the central velocities of the 2 components is noteworthy. The central 
velocity of the spectrally broad component (2) seems to agree with 
the CO one found by Winters et al. (2003), +24 \kms, and with those determined 
from the OH masers: +23.9 \kms ~for 1667 MHz and +24.6 \kms ~for 1665 MHz 
(Sivagnanam et al. 1989).

The examination of the separate position-switch data shows evidence 
of an offset West with respect to the central star.

\clearpage

\begin{figure} 
\includegraphics[angle=0,scale=.42]{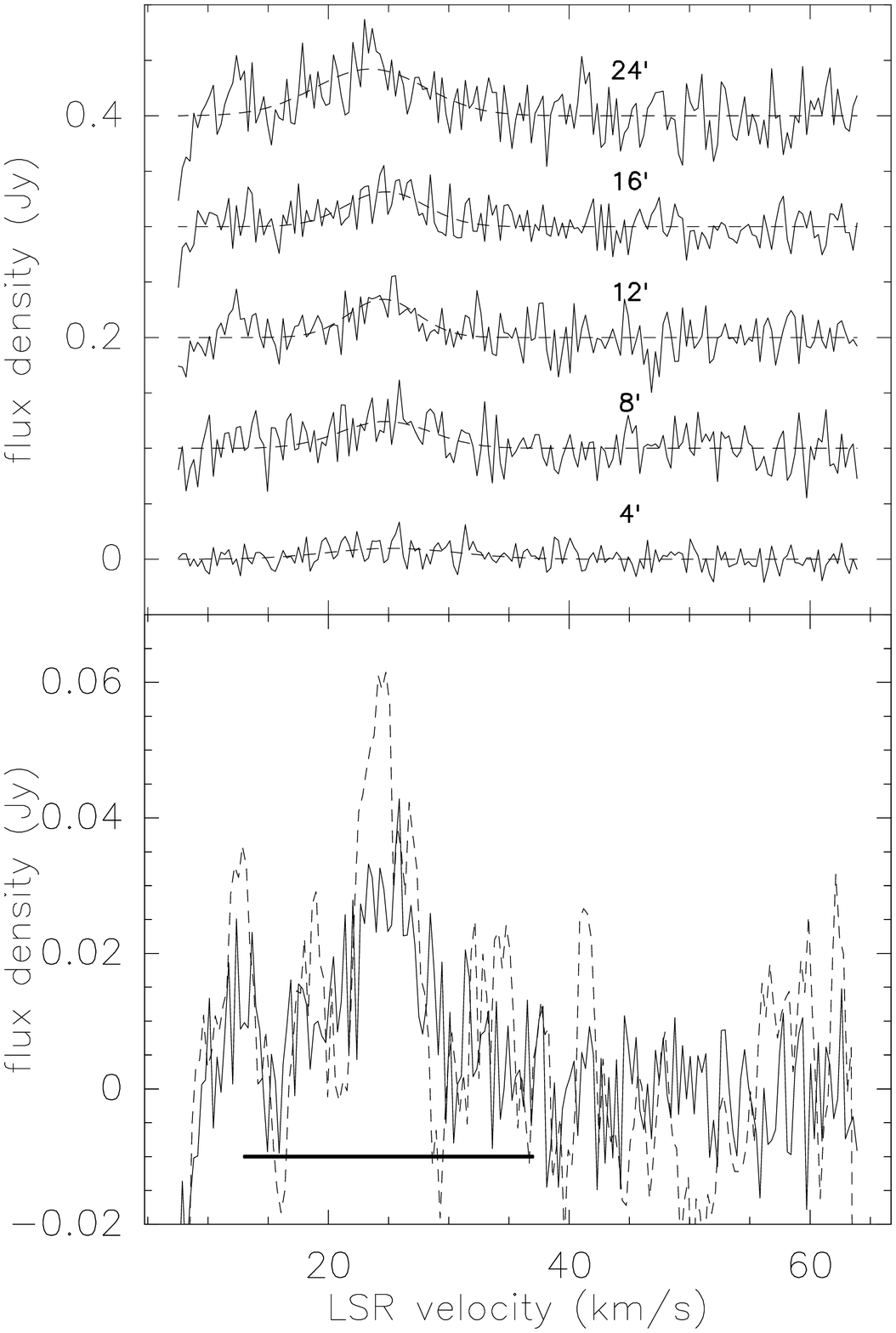}
\caption[]{Top panel: R Cas spectra obtained in the position-switch mode.
%with off-positions at $\pm 4'$, $\pm 8'$, $\pm 12'$, $\pm 16'$, and $\pm 24'$.
Bottom panel: average of these 5 spectra 
and space integrated intensity scaled by a factor 1/2.}
\label{RCas_HI_RA_av}
\end{figure}

{\bf R Cas} (HR 9066) has been observed in CO by Young (1995) and Knapp et 
al. (1998). We adopt their wind characteristics. Neri et al. (1998) 
have obtained maps in the CO (1--0) and (2--1) lines and find the source 
to be extended, $\phi \sim 17''$ and 21$''$, respectively. It is also 
associated to a 60~$\mu$m IRAS extended source of 8.6$'$ (Young et al. 1993a). 
From 2 $\mu$m interferometric data van Belle et al. (1996) determine an 
effective temperature of 2\,954 $\pm$ 174\,K, at a phase close to maximum. 

%We have obtained 33 hours of integration with off-positions at $\pm 4'$, 
%$\pm 8'$, $\pm 12'$, $\pm 16'$, and $\pm 24'$ (Fig.~\ref{RCas_HI_RA_av}).
The \HI source is barely detected at $\pm 4'$ (Fig.~\ref{RCas_HI_RA_av}), 
but clearly at $\pm 12'$. 
The diameter is thus $\sim$16$'$.

{\bf NML Tau} (IK Tau) is a large amplitude Mira whose spectral type 
%strongly changes with phase. 
exhibits important variations with phase (from M6 to M10).
It has been detected in CO (1$-$0) by 
Knapp \& Morris (1985) and Nyman et al. (1992), and mapped in the CO (2--1) 
line by Neri et al. (1998, $\phi \sim 17''$). This source presents a slightly 
self-absorbed silicate feature at 10 $\mu$m (Speck et al. 2000). 
Adopting the scale for Mira variables of 
Dyck et al. (1974), the effective temperature should vary in the range 
2\,000-3\,000\,K. We adopt the period from the infrared monitoring 
of Le~Bertre (1993), the distance from the period-luminosity relation for 
O-rich Miras (Le~Bertre \& Winters 1998) and the mass loss rate derived from 
the modelling of the infrared energy distribution (Le~Sidaner \& Le~Bertre 
1996). It was not detected in \HI by Zuckerman et al. (1980). 

\begin{figure} 
\includegraphics[angle=270,scale=.28]{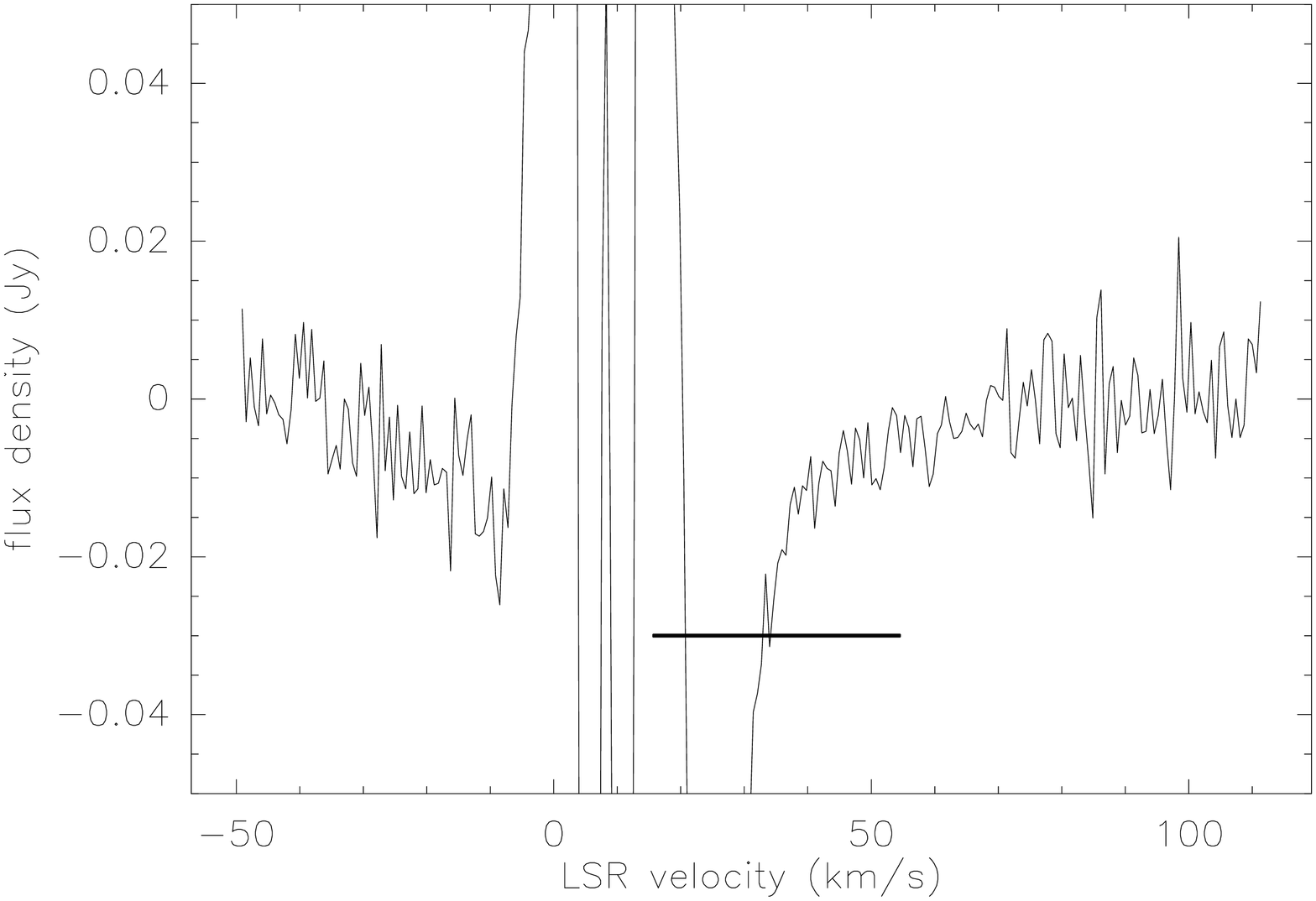} 
\caption[]{Average of all the spectra obtained in the position-switch 
mode on NML Tau. The spectral resolution is 0.64 \kms. The horizontal 
bar indicates the velocity range expected from the CO emission.
}
\label{NMLTau_HI_av}
\end{figure}

%We have obtained 31 hours of integration with off-positions at $\pm 4'$, 
%$\pm 8'$, $\pm 12'$, and $\pm 16'$. 
%Our \HI data suffer from interstellar 
%confusion over the blue half of the expected velocity range ($<$ 35 \kms), 
%but the other half is little affected except perhaps near the stellar velocity
%(Fig.~\ref{NMLTau_HI_av}). In the region $>$ 35 \kms,
% we can set an upper limit of 20 mJy to the \HI emission. Adopting 
%the CO expansion velocity and assuming that the \HI emission would be 
%symmetric with respect to 35 \kms, we find an upper limit of 
%0.8 Jy$\times$\kms~to the \HI flux within the central beam. 
Our \HI data suffer from interstellar confusion over the velocity range 
$-$40, +40 \kms. Fig.~\ref{NMLTau_HI_av} shows no convincing signal 
that could be attributed to NML Tau. Over the remaining positive velocity 
range defined by the CO line-profile (+40, +54 \kms) we find an upper limit 
to the flux density level of 20 mJy, or 0.8 Jy$\times$\kms ~over the full 
CO velocity extent of 38 \kms.
At 245 pc, this limit translates to \MHI ~$\leq$ 0.01 \Msol ~in the volume 
covered by the ``on'' position of the NRT beam. 

This limit applies to an \HI emission line. However, some circumstellar matter 
could be at a low temperature such that atomic hydrogen would appear in 
absorption against the background (see Sect.~\ref{coldHI}).
% as for IRC +10216 (Le Bertre \& G\'erard 2001). 
An upper limit to the background level is given by the sum of the 
interstellar \HI emission around +35 \kms ~($\sim$ 0.5 K) and of the continuum 
emission at 1420 MHz ($\sim$ 3.5 K, Reich \& Reich 1986), therefore 
$\sim$ 4 K. 

% In this case it is difficult to derive an \HI column density 
% because the temperature and the optical depth are correlated. 

\clearpage

{\bf WX Psc} (IRC +10011) is the most extreme oxygen-rich AGB source in 
our sample and is a representant of the type II OH/IR class, although at 
a high galactic latitude (b~=~$-50^{\circ}$). It has been 
detected in CO (1$-$0) by Knapp \& Morris (1985). The CO emission 
has been found to be extended by Neri et al. (1998, $\phi \sim 20''$).
The spectral type is probably strongly variable, as for NML Tau. 
The IRAS Low Resolution Spectrum shows a clear self-absorbed 10\,$\mu$m 
silicate feature (Volk \& Cohen 1989). The effective 
temperature should stay most of the time under 2\,500\,K (Dyck et al. 1974). 
As for NML Tau, we adopt the period from Le~Bertre (1993), the distance 
from Le~Bertre \& Winters (1998) and the mass loss rate from Le~Sidaner \& 
Le~Bertre (1996). Zuckerman et al. (1980) also attempted to detect \HI 
at 21 cm, but without success. 

%We have obtained 30 hours of integration with off-positions at $\pm 4'$, 
%$\pm 8'$, $\pm 12'$, and $\pm 16'$. 
%Our data suffer from interstellar 
%confusion at negative velocities (Fig.~\ref{WXPsc_HI_av}). 

\begin{figure} 
\includegraphics[angle=270,scale=.28]{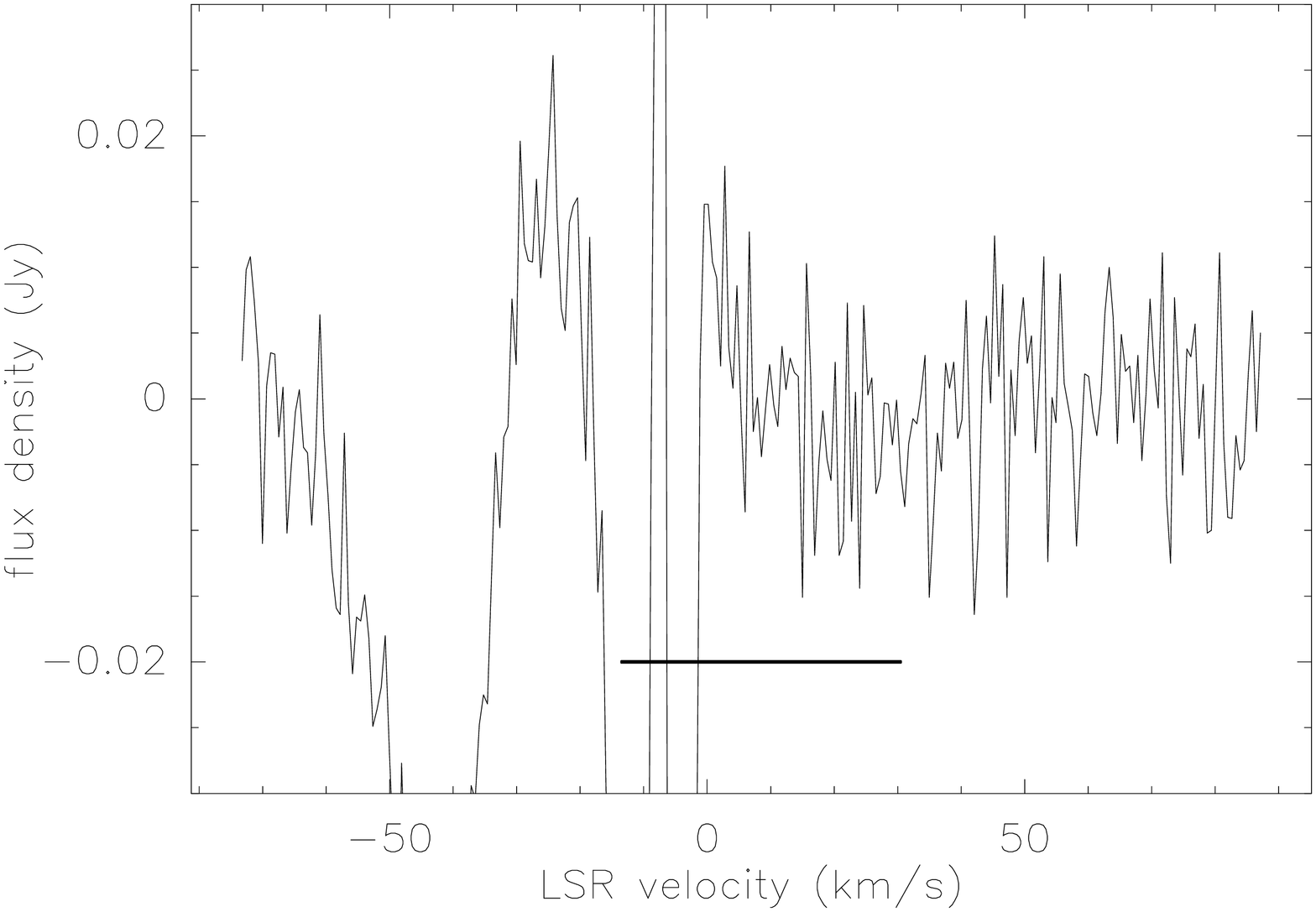}
\caption[]{Average of all the spectra obtained in the position-switch 
mode on WX Psc. The spectral resolution is 0.64 \kms. The horizontal 
bar indicates the velocity range expected from the CO emission.
}
\label{WXPsc_HI_av}
\end{figure}

Our data suffer from interstellar confusion up to +10 \kms 
~(Fig.~\ref{WXPsc_HI_av}). As for NML Tau there is no convincing signal 
that can be attributed to WX Psc.
In the region 
$>$ 10 \kms, we can set an upper limit of 5 mJy to the \HI emission. 
Following the same procedure as for NML Tau, we find an upper limit of 
0.23 Jy$\times$\kms ~to the \HI flux within the central beam. At 650 pc, 
this limit translates to \MHI ~$\leq$ 0.02 \Msol ~in the volume probed by 
the ``on'' position of the NRT beam. 

The same caveat as for NML Tau applies if this upper limit is interpreted 
as an absorption line, the background in that direction being at a level
of 4.5 K.

\clearpage

\clearpage

{
\begin{deluxetable}{lcccccccc}
\tabletypesize{\scriptsize}
\tablecaption{Gaussian fits to the \HI profiles, integrated intensities, and 
atomic hydrogen masses.\label{HItab_a}}
\tablewidth{0pt}
\tablehead{
\colhead{Source} & \colhead{confusion} & \colhead{integration} & \colhead{Figure} & \colhead{V$_{\rm cent.}$} 
& \colhead{FWHM} & \colhead{F$_{\rm peak}$} & \colhead{Flux} & \colhead{\MHI}\\
\colhead{ } & \colhead{ } & \colhead{time (hours)} & \colhead{ } & \colhead{(\kms)} & \colhead{(\kms)} 
& \colhead{(mJy)} & \colhead{(Jy$\times$\kms)} & \colhead{(\Msol)}
}
\startdata
               &           &    &                          &         &       &        &         & \\
WX Psc         & high/weak & 30 & \ref{WXPsc_HI_av}        &     ... &   ... & $<$ 5  & $<$ 0.23\tablenotemark{a} & $<$ 0.02\tablenotemark{a}\\
$o$ Cet        & weak      & 34 & \ref{oCet_HI_RA_av}      &   +45.1 &   7.7 &    73  &  0.56   &  0.0022\\
$\rho$ Per     & medium    & 50 & \ref{rhoPer_HI_RA_av}    &   +30.3 &  10.6 &   193  &  2.05   &  0.0048\\
NML Tau        & high/weak & 31 & \ref{NMLTau_HI_av}       &     ... &   ... & $<$ 20 & $<$ 0.76\tablenotemark{a} & $<$ 0.01\tablenotemark{a}\\
S CMi          & weak      & 27 & \ref{SCMi_HI_RA_av}      &   +51.0 &   5.2 &    44  &  0.23   &  0.0072 \\
U CMi          & mean      & 20 & \ref{UCMi_HI_RA_av}      &   +41.3 &   8.1 &   119  &  0.96   &  0.0083 \\
RV Hya         & weak      & 35 & \ref{RVHya_HI_av}        & $-$43.8 &   8.5 &     6  &  0.05   &  0.0012 \\
U Hya          & weak      & 42 & \ref{UHya_HI_RA_av}      & $-$30.0 &   4.9 &   102  &  0.50   &  0.0031\\
Y UMa          & weak      & 28 & \ref{YUMa_HI_RA_av}      &   +16.5 &   3.7 &    98  &  0.36   &  0.0084 \\
RY Dra         & weak      & 25 & \ref{RYDra_HI_RA_av}     &  $-$4.5 &   8.5 &   407  &  3.46   &  0.194 \\
RT Vir         & mean/weak & 24 & \ref{RTVir_HI_RA_av}     &   +13.2 &   9.6 &   192  &  1.84   &  0.0083 \\
W Hya          & high/mean & 43 & \ref{WHya_HI_RA_av}      &   +41.3 &  21.8 &   549  &  12.0   &  0.037 \\
$\alpha^1$ Her & mean/high & 46 & \ref{alphaHer_HI_RA_av}  & $-$14.4 &  11.1 &   206  &  2.29   &  0.0074 \\
NGC\,6369      & mean      & 22 & \ref{NGC6369_HI_RA_av}   & $-$91.8 &  40.5 &    31  &  1.26   &  0.71\\ 
$\delta^2$ Lyr & mean/high & 21 & \ref{delta2Lyr_HI_RA_av} & $-$10.3 &   5.7 &   737  &  4.20   &  0.075 \\
Z Cyg          & weak      & 27 & \ref{ZCyg_HI_av}         &   ...   &   ... &  $<$6  & $<$ 0.05\tablenotemark{a} & $<$ 0.003\tablenotemark{a}\\ 
NGC 7293       & weak/mean & 34 & \ref{NGC7293_HI_RA_av}   & $-$28.7 &  34.8 &   530  & 18.44   & 0.174 \\ 
R Peg          & weak      & 30 & \ref{RPeg_HI_RA_av}      &   +23.5 &   5.3 &   166  &  0.88   & 0.025 \\
AFGL 3068      & high      & 23 & \ref{AFGL3068_HI_RA_av}  & $-$26.8 &  30.9 &    80  &  2.47   & 0.76\\
AFGL 3099      & weak      & 49 & \ref{AFGL3099_HI_av}     &     ... &   ... & $<$ 2  & $<$ 0.04\tablenotemark{a} & $<$ 0.02\tablenotemark{a}\\
TX Psc         & high      & 24 & \ref{TXPsc_HI_RA_av}     &    +9.8 &   4.2 &  1250  &  5.25   & 0.067 \\ 
R Cas          & mean      & 33 & \ref{RCas_HI_RA_av}      &   +24.5 &   5.5 &   122  &  0.67   & 0.0018\\ 
\enddata
\tablenotetext{a}{in the central beam (``on-source'')}
\end{deluxetable}
}

\subsubsubsection{Carbon-rich AGB Sources}

We first discuss one Irregular and two Semi-Regular variables, then two 
high mass loss Miras, sometimes referred to as extreme carbon stars (ECS). 
As in the oxygen-rich cases, the effective temperatures of carbon Miras 
are variable and very uncertain, particularly since these two are infrared 
sources.

\clearpage

\begin{figure} 
\includegraphics[angle=0,scale=.42]{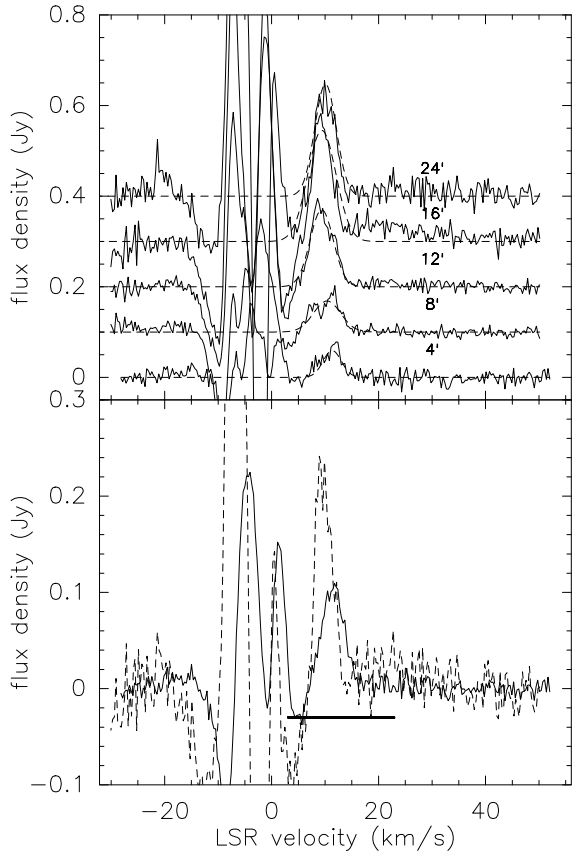}
\caption[]{Top panel: TX Psc spectra obtained in the position-switch mode. 
%with off-positions at $\pm 4'$, $\pm 8'$, $\pm 12'$, $\pm 16'$, and $\pm 24'$.
Bottom panel: average of these 5 spectra and space integrated intensity scaled 
by a factor 1/5.}
\label{TXPsc_HI_RA_av}
\end{figure}

{\bf TX Psc} (HR 9004) has a CO envelope with an irregular structure 
(Heske et al. 1989). The CO rotational line profiles may be composite 
with a narrow peak and a broad component (see also Heske 1990). 
We adopt the wind parameters from Loup et al. (1993) and the effective 
temperature (3\,115 $\pm$ 130 K) from Bergeat et al. (2001). The source 
has been found to be extended at 60 $\mu$m by IRAS (Young et al. 1993a).

%We have obtained 24 hours of integration with off-positions at $\pm 4'$, 
%$\pm 8'$, $\pm 12'$, $\pm 16'$, and $\pm 24'$ (Fig.~\ref{TXPsc_HI_RA_av}).
The \HI signal grows rapidly with increasing offset 
(Fig.~\ref{TXPsc_HI_RA_av}). We estimate the \HI diameter at 24$'$. 

\clearpage

\begin{figure} 
\includegraphics[angle=0,scale=.42]{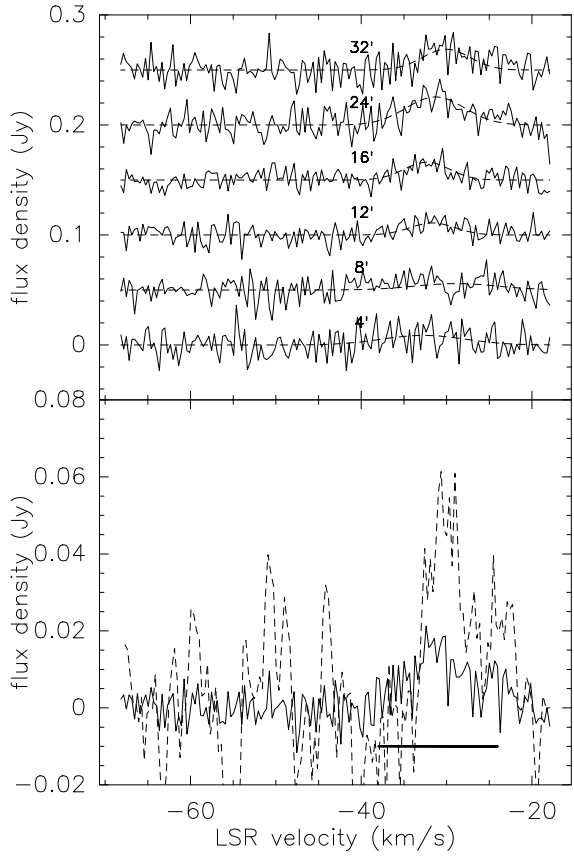}
\caption[]{Top panel: U Hya spectra obtained in the position-switch mode. 
%with off-positions at $\pm 4'$, $\pm 8'$, $\pm 12'$, $\pm 16'$, $\pm 24'$, 
%and $\pm 32'$. 
Bottom panel: average of these 6 spectra and space integrated 
intensity scaled by a factor 1/2.}
\label{UHya_HI_RA_av}
\end{figure}

{\bf U Hya} (HR 4163) has been detected in CO (2$-$1) by Knapp et al. (1998) 
and we adopt their wind parameters. 
The IRAS data show a detached dust shell ($\phi \sim 3.5'$) 
produced by a previous episode of much higher mass loss rate, 
$\sim$ 5$\times$10$^{-6}$ \Msold, (Waters et al. 1994). 
We take the effective temperature from Bergeat et al. (2001).

%We have obtained 42 hours of integration with off-positions at $\pm 4'$, 
%$\pm 8'$, $\pm 12'$, $\pm 16'$, $\pm 24'$, and $\pm 32'$ 
%(Fig.~\ref{UHya_HI_RA_av}). 
The \HI line is faint (Fig.~\ref{UHya_HI_RA_av}), but well detected 
on the average spectrum, and might be composite. 
Also the emisson is spatially resolved ($\phi \sim 32'$). Although the flux 
is weak, this case might be similar to Y CVn.

{\bf RY Dra} belongs to the rare J-type class, like the star Y CVn 
which was discussed in Paper II. It has been detected in the $^{12}$CO 
and $^{13}$CO rotational lines by Jura et al. (1988), the latter with 
about half the intensity of the former. The ISO data show a slight evidence 
of a detached shell that Izumiura \& Hashimoto (1999) ascribe to 
a previous episode of mass loss at a rate of $\sim$ 10$^{-6}$ \Msold. 
Furthermore, the IRAS data show a very extended source ($\phi \sim$ 40$'$) 
at 60 and 100 $\mu$m (Young et al. 1993a). We adopt the effective temperature 
from Bergeat et al. (2001).

\begin{figure} 
\includegraphics[angle=0,scale=.42]{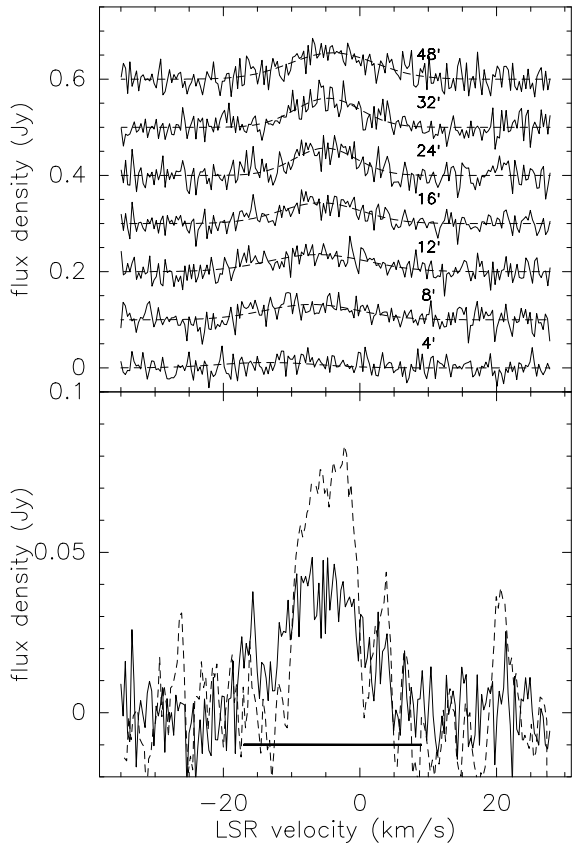}
\caption[]{Top panel: RY Dra spectra obtained in the position-switch mode.
%with off-positions at $\pm 4'$, $\pm 8'$, $\pm 12'$, $\pm 16'$, 
%$\pm 24'$, $\pm 32'$, and $\pm 48'$. 
Bottom panel: average of these 
7 spectra and space integrated intensity scaled by a factor 1/5.}
\label{RYDra_HI_RA_av}
\end{figure}

%We have obtained 25 hours of integration with off-positions at $\pm 4'$, 
%$\pm 8'$, $\pm 12'$, $\pm 16'$, $\pm 24'$, $\pm 32'$, and $\pm 48'$ 
%(Fig.~\ref{RYDra_HI_RA_av}). 
The \HI source is weakly detected at small offsets (Fig.~\ref{RYDra_HI_RA_av})
but reaches more than 50 mJy at $\pm 24'$. The diameter is the largest 
in our sample, $\sim$ 40$'$. 

\clearpage

The 2 following sources were discovered in the AFGL infrared survey.
Their carbon-rich nature was established from near-infrared spectrophotometry
by the presence of the 3.1~$\mu$m carbon-star absorption feature 
(Gehrz et al. 1978; Jones et al. 1978).

\clearpage

\begin{figure} 
\includegraphics[angle=270,scale=.28]{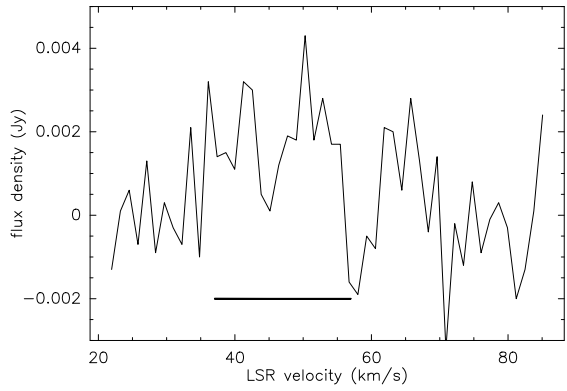} 
\caption[]{Average of all the spectra obtained in the position-switch 
mode on AFGL\,3099. The spectral resolution is 1.28 \kms. The horizontal 
bar indicates the velocity range expected from the CO emission.
}
\label{AFGL3099_HI_av}
\end{figure}

{\bf AFGL 3099} (IZ Peg) was detected in CO (1$-$0) by Knapp \& Morris 
(1985). The CO (2$-$1) emission has been mapped by Neri et al. (1998) 
and shown to be asymmetric (14$''\times$7$''$). We adopt 
the period from Le Bertre (1992), the distance from the period-luminosity 
relationship for carbon Miras (Groenewegen \& Whitelock 1996)  
and the mass loss rate from the modelling of the infrared energy distribution 
(Le Bertre 1997). There is no stellar spectral classification; however, as 
the 3.1~$\mu$m feature is extremely deep (Gehrz et al. 1978), which is an 
indication of a very low temperature, we adopt the effective temperature 
for the coolest carbon stars in Bergeat et al. (2001).

%We have obtained a total of 49 hours of integration with off-positions at 
%$\pm 4'$, $\pm 8'$, $\pm 12'$ and $\pm 24'$. 
We could not detect the source for any beam-throw (up to $\pm$ 6 beams, 
i.e. $\pm 24'$). 
In Fig.~\ref{AFGL3099_HI_av} we present the average of all 
our \HI spectra. This is the deepest integration in our survey.
%and we may have a hint for a detection at the expected radial velocity. 
In the 30--80 \kms ~range the \HI interstellar background is low 
and perfectly removed by our procedure, so that we can safely set an upper 
limit of 2$\times$10$^{-3}$ Jy to any circumstellar \HI emission feature in 
the central position. Adopting the CO expansion velocity, we find an upper 
limit of 0.04 Jy$\times$\kms ~to the \HI emission, which, at a distance 
of 1.5 kpc, translates to \MHI ~$\leq$ 2$\times$10$^{-2}$ \Msol ~in 
the volume probed by the on-position.

{\bf AFGL 3068} (LL Peg) has such a dense wind that the SiC dust feature at 
11 $\mu$m is seen in absorption (Jones et al. 1978). It has been detected in 
CO (1$-$0) by Knapp \& Morris (1985) and Nyman et al. (1992). The CO (2$-$1) 
emission has been mapped by Neri et al. (1998) who suggest the presence 
of a large ($\phi \sim 17''$) detached bipolar shell. Mauron \& Huggins 
(2006) have detected galactic light scattered by circumstellar dust out 
to a distance of $\sim$ 40$''$ from the central star with an amazing spiral  
structure. We adopt the period and the distance from Le Bertre (1992, 1997). 
A mass loss rate of 5$\times$10$^{-5}$ \Msold ~was derived from a radiative 
transfer model of the infrared energy distributions by Le Bertre et al. (1995) 
whereas a consistent time-dependent hydrodynamical model of this source 
obtained by Winters et al. (1997) gives $\sim$ 1$\times$10$^{-4}$ \Msold.  
As for AFGL\,3099, we adopt the effective temperature for the coolest carbon 
stars (Bergeat et al. 2001).

\begin{figure} 
\includegraphics[angle=0,scale=.42]{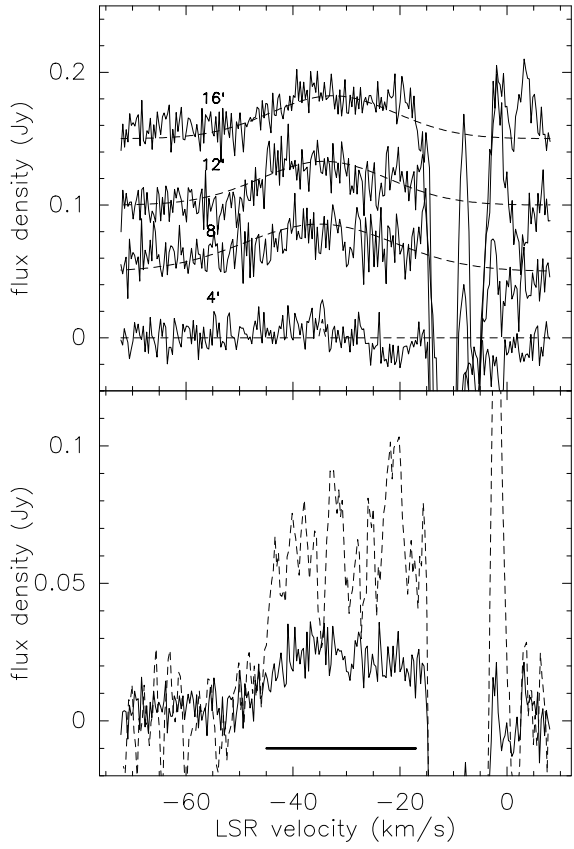}
\caption[]{Top panel: AFGL\,3068 spectra obtained in the position-switch mode.
%with off-positions at $\pm 4'$, $\pm 8'$, $\pm 12'$, and $\pm 16'$. 
Bottom panel: average of these 4 spectra and space integrated intensity.}
\label{AFGL3068_HI_RA_av}
\end{figure}

%Position-switch data have been acquired at $\pm 1$, $\pm 2$, $\pm 3$, and 
%$\pm 4$ beams (23 hours). 
An emission feature centered at $-$31~\kms ~is 
barely present at $\pm 4'$, but clearly detected at $\pm 8'$, $\pm 12'$ 
and $\pm 16'$ (Fig.~\ref{AFGL3068_HI_RA_av}). 
The intensity is increasing from $\pm 4'$ to $\pm 8'$, 
but not significantly from $\pm 8'$ to $\pm 16'$. 
We conclude that \HI is detected 
in emission from AFGL\,3068 and that this emission is extended with a size 
of at least $\sim$ 8$'$ in diameter, or 2.6 pc at 1.14 kpc.

\clearpage
 
\subsubsection{PN Sources}\label{Sample_PN}

After they have expelled most of their stellar envelope red giants 
are thought to evolve toward the Planetary Nebula (PN) stage. 
The mass loss decreases and the hot central core progressively emerges 
and starts to ionize its surrounding. For some time, a neutral remnant from 
the AGB circumstellar shell can still be present around an ionized nebula. 
Recent developments may imply that this scenario applies only to the 
giants with more massive progenitors, and/or to those that are in a close 
binary system and experience a common envelope interaction.

``Circumnebular'' neutral hydrogen has been detected in absorption against 
the continuum emitted by the central ionized nebula in 5 PNs by Taylor 
et al. (1990). Atomic hydrogen has also been detected in emission from 
IC 418 (Taylor et al. 1989) and from NGC 7293 (R2002).

{\bf NGC\,6369} (Little Ghost Nebula) is a PN of optical diameter $\sim 40''$. 
The central star (HD\,158269) is an hydrogen-deficient [WC] Wolf-Rayet 
star. A detailed photoionization model of the nebula by Monteiro et al. (2004) 
shows that it has a hourglass structure. They derive a nebular mass of 
1.8 \Msol. We adopt the distance derived from this modelling and 
the expansion velocity (\Vexp = 41.5 \kms)
determined from the [{\sc O\,iii}] line width by 
Meatheringham et al. (1988). We note that the [{\sc O\,iii}] line should probe 
the ionized central part of the nebula, and that it probably overestimates 
the expansion velocity of the external medium. Indeed, Garay et al. (1989)
have obtained an H76${\alpha}$ linewidth of only 42.8 \kms.
This source has been seen to be extended by IRAS 
at 25 $\mu$m ($\phi \sim 1.1'$); this extension could be related to 
the un-identified 21-$\mu$m emission feature reported by Hony et al. (2001).

\begin{figure} 
\includegraphics[angle=0,scale=.42]{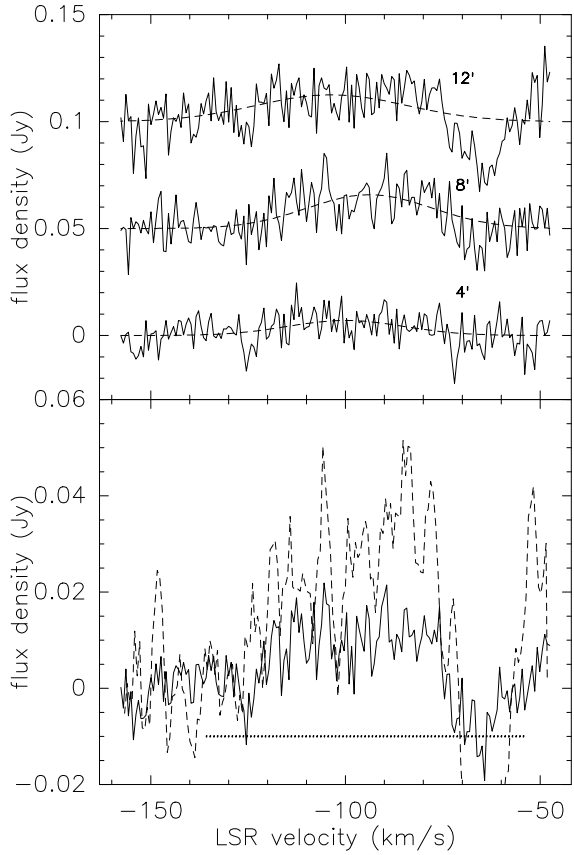}
\caption[]{Top panel: NGC\,6369 spectra obtained in the position-switch mode. 
%with off-positions at $\pm 4'$, $\pm 8'$, and $\pm 12'$. 
The spectral resolution is 0.64 \kms. Bottom panel: average of these 3 spectra 
and space integrated intensity. The horizontal dashed bar indicates 
the velocity range expected from the [{\sc O\,iii}] line-width 
(Meatheringham et al. 1998).}
\label{NGC6369_HI_RA_av}
\end{figure}

%We have obtained 22 hours of integration with off-positions at $\pm 4'$, 
%$\pm 8'$, and $\pm 12'$ (Fig.~\ref{NGC6369_HI_RA_av}).  
The source is barely detected at $\pm 4'$ (Fig.~\ref{NGC6369_HI_RA_av}, 
but clearly at $\pm 8'$. We estimate its diameter at 8$'$.

\clearpage

{\bf NGC 7293} (Helix Nebula) is a nearby PN ($\sim$\,200\,pc) with a large 
extent on the sky ($\phi \sim$\,1000$''$). Young et al. (1999) have mapped 
the nebula in the CO (2$-$1) line. The CO linewidth is about 63 \kms.
The emission is localized in condensations tracing an expanding equatorial 
ring that is seen almost pole-on and that surrounds the bright ionized 
nebula (Huggins et al. 1999). 
Speck et al. (2002) have obtained H$_2$ (v=1$-$0) line map which shows that 
this emission also coincides with the bright nebula.  
Furthermore they present an H$\alpha$ image which shows a weak and more 
extended emission, up to $\sim$\,1100$''$ from the central star. 
This weak emission coincides with dust emission detected by IRAS at 60 
and 100 $\mu$m, and by ISO at 160 $\mu$m. 
Neutral carbon ({\sc C\,i} line at 492 GHz) was detected in several positions 
on the bright nebula by Young et al. (1997). 

NGC 7293 was observed in \HI with the VLA (R2002). The emission is 
fragmented and follows the CO (2$-$1) ring, except in the South-East quadrant 
where it is missing. With a diameter of $\sim 16'$, 
it surrounds the bright optical nebula; in particular 
it is absent in the central part ($\phi \sim$\,500$''$). R2002 estimate 
the \HI mass to $\sim$ 0.07 \Msol. However, from the continuum data 
at 21 cm, they estimate that they may have missed $\sim$ 50\,\% (0.67/1.4) 
of the flux. 

\clearpage

\begin{figure} 
\includegraphics[angle=0,scale=.42]{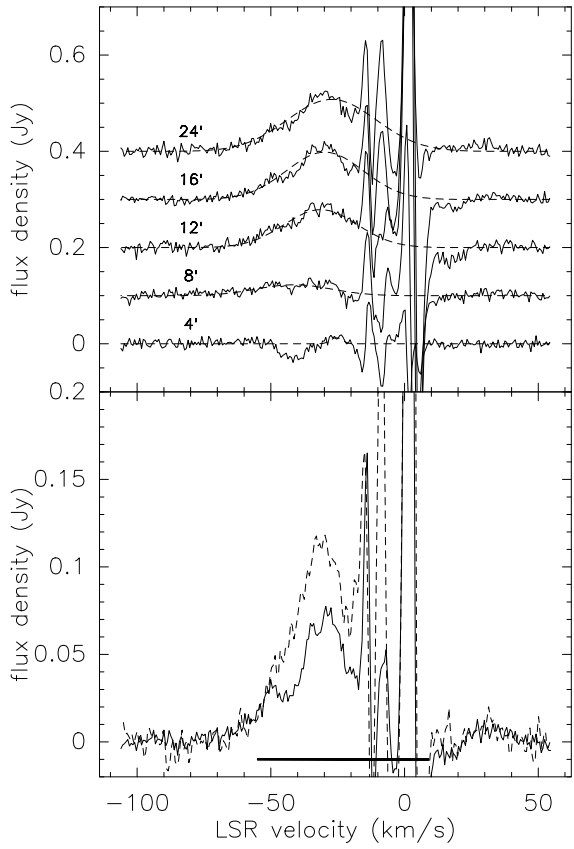}
\caption[]{Top panel: NGC\,7293 spectra obtained in the position-switch 
mode. 
%with off-positions at $\pm 4'$, $\pm 8'$, $\pm 12'$, $\pm 16'$, and $\pm 24'$.
The spectral resolution is 0.64 \kms. 
Bottom panel: average of these 5 spectra, and space integrated 
intensity scaled by a factor 1/5.}
\label{NGC7293_HI_RA_av}
\end{figure}

%We have obtained 34 hours of integration with off-positions at $\pm 4'$, 
%$\pm 8'$, $\pm 12'$, $\pm 16'$, and $\pm 24'$ (Fig.~\ref{NGC7293_HI_RA_av}).
The interstellar contamination seems large at V $> -$20 \kms. The source 
is clearly detected in emission for V $< -$20 \kms 
~(Fig.~\ref{NGC7293_HI_RA_av}) and throws larger than 
8$'$. We detect nearly the same intensity at $\pm 16'$ and $\pm 24'$, 
and thus evaluate the diameter at $\sim 24'$. This is much larger than 
estimated by R2002, but consistent with the finding of H$\alpha$ and 
dust emissions up to 18$'$ from the central star by Speck et al. (2002).

A comparison of our \HI spectrum with the spatially integrated spectrum of 
R2002 shows that the narrow feature at $-$15 \kms ~probably belongs 
to the source and is not due to confusion, i.e. the contamination may 
affect our data only at V $> -$15 \kms. Also, the intensity of this feature 
levels off at $\pm$ 12$'$, an indication that it is not due to the 
galactic background. Nevertheless we have derived \HI line parameters 
(Table \ref{HItab_a}) by using only the part of the spectrum outside the range 
($-$20, +20 \kms). We find a central velocity $\sim -$29 \kms, instead 
of $-$24 \kms ~(Young et al. 1999), probably because we excluded the $-$15 
\kms ~feature from the fit. Also we observe in our average spectrum a narrow 
component at $-$50 \kms ~that is not seen in the R2002's spectrum. 
However, a CO feature is detected at this velocity by Young et al. (1999). 
The latter is concentrated in the South-East part of the Helix ring where 
R2002 detect no \HI emission.

For the main component centered at $-$29 \kms, the 
examination of the separate position-switch data does not show evidence 
of a spatial offset with respect to the central star. On the other hand 
the emission at $-$15 \kms ~is offset to the West.
%in agreement with R2002 who locate it to the North-West. 
R2002 have mapped this emission with the VLA (their figure 7). 
The corresponding channel map shows that the feature is located in 
the North-West and that it could be due to an interaction with the local ISM.

For the main component, we measure an 
integrated \HI intensity of $\sim$ 18.4 Jy$\times$\kms, which is a factor 2.3 
larger than the R2002's estimate. This is in agreement with their finding 
that, with the VLA, they may have lost $\sim$ 50\% of the flux. The \HI 
emission missed by R2002 is probably diffuse and extends beyond the classical 
torus of diameter 16$'$. The intensity that we missed in 
the narrow feature at $-$15 \kms ~is only 0.5~Jy$\times$\kms. 
Considering the extent that we find and the 
North-South size of our beam (22$'$), it is likely that we also missed 
\HI flux, at the level of $\sim$ 30 \%, to the North and to the South of 
the central source, bringing the total flux to $\sim$ 28~Jy$\times$\kms. 
A full map similar to those that we did in Paper II for EP~Aqr and Y~CVn 
would be useful. 

The \HI flux from NGC 7293 is indeed so large that the source can be seen 
on the Hartmann \& Burton (1997) maps obtained at velocities close to 
$-$40 \kms ~for which the galactic confusion is moderate.

\section{INTERPRETATION}\label{interpretation}

\subsection{Line Profiles}\label{lineprofiles}

One of the most striking results emerging from our survey is that 
we find \HI line-profiles with a quasi-Gaussian shape, whereas
we were expecting rectangular profiles, for unresolved sources, or 
double-horn ones, for resolved sources. 
Sometimes the profile seems 
triangular with a peak close to the centroid velocity (e.g. RT Vir, 
Fig.~\ref{RTVir_HI_RA_av}, or $o$ Cet, Fig.~\ref{oCet_HI_RA_av}).

In Paper II, we indeed performed numerical simulations of the \HI emission 
produced by a spherical circumstellar envelope. The emission is assumed 
to remain optically thin, and its brightness temperature to be proportional 
to the \HI column density. Therefore there is no shell self-absorption 
and the velocities are radial.
In these conditions for an unresolved source 
with a uniform wind velocity and a constant production rate in atomic 
hydrogen, the line-profile is rectangular, is centered on the stellar radial 
velocity, and has a width equal to twice the expansion velocity. For 
a resolved source, the line profile shows two horns with peaks separated 
by twice the expansion velocity. These simulations were performed assuming 
a uniform response within the telescope beam. However, introducing a more 
realistic response did not significantly change the results (see e.g. figure 7 
in Paper III).

We seldom observe double-horn profiles. Only $\alpha$\,Her 
may show such a shape. We also note that NGC\,7293 
shows several components that could be interpreted as horns. 
Likewise, the profiles of AFGL\,3068 and NGC\,6369 could be non-Gaussian.  
Previously, in a few peculiar cases, we detected a component of 
the \HI emission that might be fitted by a rectangle (RS Cnc, 
Paper I; EP Aqr, Paper II; X Her, Paper III). In these cases, the width is 
relatively narrow, $\sim$ 3-4 \kms. 

To get centrally peaked profiles we had 
to assume that the \HI velocity is varying with distance from the central 
star. For resolved sources the velocity has to decrease to zero relative to 
the stellar radial velocity. A large quantity of hydrogen has to accumulate 
close to zero velocity, otherwise a deficit appears 
in the center of the profile. For unresolved sources, the velocity 
could be assumed to decrease as well as to increase with distance. 
However, it would be unlikely to get increasing 
velocities only for unresolved sources, and therefore we can conclude that, 
on scales probed by the \HI emission, the expansion velocity always decreases 
with distance from the central star. This slowing-down of the circumstellar 
wind could be due to its interaction with external matter, either remains 
of previous 
episodes of mass loss and/or surrounding ISM (Young et al. 1993b; Paper II).

\clearpage

\begin{figure} 
\includegraphics[angle=-90,scale=.3]{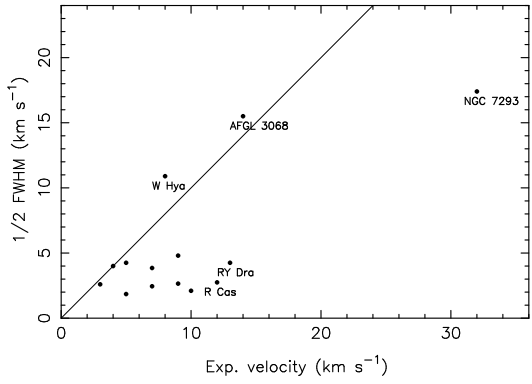}
\caption[]{Half-FWHM of the \HI line versus expansion velocity determined 
from CO rotational lines.}
\label{vitesse}
\end{figure}

\clearpage

The slowing-down is also suggested by the comparison of the expansion 
velocity derived from the CO line-profiles to the half-FWHM of the \HI line
(Fig.~\ref{vitesse}). Except for 
AFGL 3068 and W Hya, the former is always larger than the latter. We note that 
AFGL 3068 is at a large distance from the galactic plane ($\sim$ 740 pc),  
perhaps in a region devoid of interstellar matter.

In general a gaussian profile can result from thermal and turbulent 
broadening, as well as from systematic motions of the gas. 
Presently in our modelling, we are not taking into account 
thermal and turbulent effects. On the one hand
their inclusion in the model may force to reduce the \HI ~outflow velocity
because they will provide a supplementary source of line broadening.
On the other hand the velocity gradient may have to be reduced and outflow
velocities close to 0 (relative to the central star) may not be needed
(Libert, G\'erard \& Le Bertre, in preparation).

\subsection{Spatial Distributions}\label{spatialdistributions}

\clearpage

{
\begin{deluxetable}{lcccccccc}
\tabletypesize{\scriptsize}
\tablecaption{Estimated characteristics: sizes, timescales, 
average mass loss rates.\label{Estimates_tab}}
\tablewidth{0pt}
\tablehead{
\colhead{Source} & \colhead{Chemistry} & \colhead{$\phi_{H I}$} & Total \HI mass & \colhead{diameter} 
& \colhead{$\tau$} & \colhead{$<${\.M}$>$} & \colhead{$\phi_{IRAS}$}\\
\colhead{ } &      & \colhead{($'$)} & (\Msol) & \colhead{(pc)} & \colhead{(10$^3$ yr)} 
& \colhead{(\Msold)} & \colhead{($'$)}
}
\startdata
               &      &     &          &        &      &                      &      \\
$o$ Cet        &  O   &  8. &  0.0023  &  0.30  &  38  & 0.8$\times$10$^{-7}$ &  4.4 \\
$\rho$ Per     &  O   & 24. &  0.0071  &  0.70  &  64  & 1.5$\times$10$^{-7}$ &  ... \\
S CMi          &  O   &  8. &  0.0076  &  0.85  & 160  & 0.6$\times$10$^{-7}$ &  ... \\
U CMi          &  O   &  8. &  0.0088  &  0.44  &  54  & 2.2$\times$10$^{-7}$ &  ... \\
RV Hya         &  O   &  2. &  0.0012  &  0.19  &  21  & 0.8$\times$10$^{-7}$ &  ... \\
U Hya          &  C   & 32. &  0.0055  &  1.51  & 301  & 0.2$\times$10$^{-7}$ &  5.8 \\
Y UMa          &  O   &  8. &  0.0089  &  0.73  & 193  & 0.6$\times$10$^{-7}$ &  7.6 \\
RY Dra         &  C   & 40. &  0.403   &  5.68  & 654  & 8.2$\times$10$^{-7}$ & 37.8\tablenotemark{a}\\
RT Vir         &  O   & 24. &  0.0123  &  0.96  &  98  & 1.7$\times$10$^{-7}$ &  8.2 \\
W Hya          &  O   & 24. &  0.055   &  0.80  &  36  & 2.0$\times$10$^{-6}$ & 21.0\tablenotemark{b}\\
$\alpha^1$ Her &  O   & 16. &  0.0091  &  0.54  &  48  & 2.5$\times$10$^{-7}$ &  ... \\
NGC 6369       &  O   &  8. &  0.76    &  3.6   &  87  & 1.2$\times$10$^{-5}$ &  ... \\
$\delta^2$ Lyr &  O   & 24. &  0.11    &  1.92  & 330  & 4.5$\times$10$^{-7}$ & 33.0\tablenotemark{c}\\
NGC 7293       &  O   & 24. &  0.258   &  1.40  &  39  & 8.7$\times$10$^{-6}$ &  ... \\
R Peg          &  O   & 16. &  0.031   &  1.63  & 301  & 1.4$\times$10$^{-7}$ &  ... \\
AFGL 3068      &  C   &  8. &  0.81    &  2.65  &  84  & 1.3$\times$10$^{-5}$ &  ... \\ 
TX Psc         &  C   & 24. &  0.100   &  1.63  & 380  & 3.5$\times$10$^{-7}$ &  6.2 \\
R Cas          &  O   & 16. &  0.0022  &  0.50  &  89  & 0.3$\times$10$^{-7}$ &  8.6 \\
\enddata
\tablenotetext{a}{44.2$'$ at 100 $\mu$m}
\tablenotetext{b}{30.0$'$ at 100 $\mu$m}
\tablenotetext{c}{33.8$'$ at 100 $\mu$m}

\end{deluxetable}
}

Although the spatial resolution of the NRT is low, most of our sources are 
so large that they are resolved in the East-West direction. In Table 
\ref{Estimates_tab} we give estimates of the diameters of the \HI sources 
(col. 3) as explained in Sect.~\ref{obs}.  
For sources that are not resolved with the NRT, we adopt a diameter of 2$'$.
The four sources that were not detected in \HI emission (WX Psc, NML Tau, 
Z Cyg and AFGL 3099) have not been reported in this Table.
We then correct (col. 4) the measured \HI mass assuming that the \HI emission 
is circularly symmetric, i.e. it has the same size in declination and 
in right ascension. 
The sizes of the \HI shells (col. 5) are given for the distances adopted 
in Table~\ref{Sample_tab}. A characteristic timescale (col. 6) can be 
derived from this size and the FWHM of the \HI emission line. Finally from 
the estimated mass of hydrogen in the circumstellar shell, and assuming 
that 75 $\%$ of the mass is in hydrogen, we derive an average  mass loss rate 
over this timescale (col. 7). 

\clearpage

\begin{figure} 
\includegraphics[angle=-90,scale=.3]{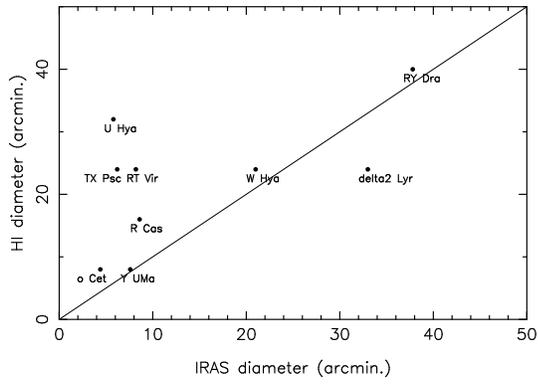}
\caption[]{\HI versus IRAS 60 $\mu$m diameter from Young et al. (1993a).}
\label{diametre}
\end{figure}

\clearpage

Young et al. (1993a) measured the angular extent of red giants and young PNs  
in the IRAS 60 and 100 $\mu$m survey data. For comparison, we give also the 
diameter estimated by these authors from the 60 $\mu$m IRAS data (col. 8). 
In general we find sizes in \HI that are comparable or larger 
(Fig.~\ref{diametre}). This should not be overemphasized as \HI diameters 
are only rough estimates.

$o$ Cet was barely resolved with the NRT, although it is 
one of the closest sources in our sample. We find 
an estimate of the size of the \HI source (in the East-West direction) 
which is consistent with the size of the dust shell estimated from the IRAS 
data at 60 $\mu$m (R $\sim 2.2'$, or 0.08 pc, Young et al. 1993a). 
We note that BK1988 detected \HI emission up to $\sim$ 160$''$ (2.7$'$) from 
the central star. At the other extreme, AFGL\,3068, which is at $\sim$ 
1.2 kpc, was angularly resolved with the NRT, $\phi \sim 8'$, or $\geq 2$ pc. 
The \HI circumstellar shells of Miras may therefore have very different 
physical sizes, suggesting very different durations for the past mass loss. 

We also note that in the case of AFGL\,3068 the \HI shell seems much larger 
than the shell observed in the far-infrared that was not resolved 
by IRAS (Young et al. 1993a) or by ISO (Izumiura \& Hashimoto 1999). 
This may be an effect of the dust emission dependence on temperature 
that may impede the detection of matter far from the central star or 
indicate a relatively recent onset of copious mass-loss rate that 
is not reflected in the HI spectra. Interestingly, our \HI data imply 
a mass-loss rate about 10 times lower than that given by the detailed 
modelling of Winters et al. (1997).

ISO observations at 60 and 90 $\mu$m of WX Psc show a point source with 
a FWHM $\sim 1'$ (Hashimoto \& Izumiura 2000). For a distance of 650 pc, 
the circumstellar shell diameter should be at most 0.2 pc. This is interpreted 
as a recent evolution with a sudden event ($<$ 4000 years) of extremely  
high mass-loss rate. However, Hashimoto \& Izumiura do not exclude the 
possibility of a faint extended emission that would be the trace of a past 
mild mass-loss phase. A recent onset of the mass loss in sources like WX Psc 
or NML Tau (that was not resolved by IRAS, Young 1993a) might explain their 
non-detection in H\,{\sc {i}}. Indeed in young compact sources with central 
star effective temperature lower than 2500\,K, most of the circumstellar 
matter may be molecular (see Sect.~\ref{molhydrogen}).

We have also examined our position-switch data by using separately the two 
off-positions. The profiles are severely affected by the spatial variations 
of the background and the resulting spectra suffer much more from galactic 
\HI confusion, which makes them in general difficult to exploit. 
%Nevertheless, 
In some cases (weak/mean confusion, large signal), e.g. RT Vir, U CMi, 
we can detect the circumstellar emission and observe that it is not 
symmetric with respect to the central position (i.e. the star position).
Nevertheless, this should be taken with caution as 
the increasing confusion towards 0 \kms ~velocity may 
affect the baselines of the separate East and West profiles. However, 
the same effect was noted previously for EP Aqr (Paper II) and X Her 
(Paper III) for which we also observed shifts in the North-South direction.
Finally we note that in the $o$~Ceti map obtained with the VLA by BK1988 
the \HI emission appears offset to the North-West.

From IRAS data at 60 $\mu$m Young et al. (1993b) find that among nearby stars 
Semi-Regular variables are more likely to be resolved than Miras. 
Our statistics is small but our results (at least for oxygen-rich sources)
tend to agree with their conclusion that Semi-Regular variables 
have been losing mass for a longer duration than Miras. On the other hand 
they do not find evidence that the shapes of the shells are distorted by 
interaction with the ISM, which we do (see also Sect.~\ref{interaction_ISM}).
Finally, AGB carbon-rich sources appear more extended than oxygen-rich 
ones. The difference is still more striking when comparing the total \HI 
masses (col. 3 in Table~3). This quantity ranges from 0.001 to 0.1 \Msol 
~for oxygen-rich sources, and from 0.005 to 0.7 \Msol ~for carbon-rich 
sources. 

The larger extension of carbon-rich sources as compared to oxygen-rich 
ones is not so surprising as carbon-rich AGB stars should be in a phase 
of evolution that follows the oxygen-rich stage. And the longer the star 
loses mass, the larger the mass of its circumstellar shell.

\subsection{\HI versus CO}\label{HIversusCO}

We have detected \HI emission in sources with and without emission from the 
CO rotational lines. For instance $\alpha^1$ Her is detected with a peak 
intensity $\sim$ 50 mJy at $\pm 12'$, but was not detected in CO by Heske 
(1990) although the presence of circumstellar matter close to the central star 
was demonstrated by Deutsch (1956). We note that this situation appears 
preferentially for warm giants with \Teff $>$ 3\,000 K. It suggests that CO 
is underabundant in these sources for which the outflows might be mostly 
atomic. If this suggestion is correct then these sources could be 
detectable in O\,{\sc {i}} ~and C\,{\sc {i}} ~lines, for instance at 492 GHz. 
%Unfortunately 
Knapp et al. (2000) did not detect C\,{\sc {i}} ~at 492 GHz from 
$\alpha^1$ Her, but this could be an effect of a cyclic variation 
of the mass loss rate, the region that they probed being only 15$''$ 
in diameter. We also note that our estimate for the average mass loss rate 
is 2.5 times larger than that of Deutsch (1956).

We sometimes easily detect \HI from Semi-Regular variables with low mass loss 
rates whereas we may not detect it from Miras with large mass loss rates. 
This may be an effect of the dependence of the H$_2$/\HI equilibrium 
in the stellar atmosphere on temperature. Indeed the stellar effective 
temperature of the former are larger than 2\,500 K, and smaller for 
the latter, and GH1983 show that hydrogen should be mostly atomic 
for \Teff $>$ 2\,500 K, and molecular for \Teff $<$ 2\,500 K. Atomic 
hydrogen should then be present only in the external parts of the 
circumstellar shells of Miras with large mass loss rates where 
H$_2$ is photo-dissociated by the Interstellar Radiation Field. This point 
is further discussed in Sect.~\ref{coldHI} and \ref{molhydrogen}.
CO and \HI can trace outflows only where these species are present.

If we now compare CO and \HI profiles of sources that show both emissions, 
we find that, with the notable exception of $o$ Cet, they often differ. 
CO profiles are generally parabolic or rectangular whereas \HI profiles 
are Gaussian or triangular. The \HI emission is found within the range 
encompassed by the CO emission.
%, but is generally narrower (Fig.~\ref{vitesse}).
The \HI FWHM is generally smaller than the expansion velocity determined 
from CO data (Fig.~\ref{vitesse}).
Sometimes its centroid is shifted by 1-3 \kms ~with respect to the CO 
centroid. For instance, in Y~UMa, we find \HI emission from 14 to 20 \kms 
~with a centroid at 17 \kms, whereas the CO emission spans from 14 to 24 
\kms ~and is centered at 19 \kms ~(Knapp et al. 1998). Although the 
kinematics of the outflows should be the same in CO and H\,{\sc {i}}, the 
regions that are probed by these two tracers are different. Firstly, the beams 
are different, typically 10-50$''$ for CO observations, and 4$'\times$22$'$ 
for our \HI observations. Secondly, CO and atomic hydrogen may be present 
in different zones of the circumstellar shells. For instance CO is 
expected to be photodissociated by the interstellar radiation field 
at distances $\sim$ 10$^{16}$-10$^{18}$ cm 
depending on the mass loss rate (Mamon et al. 1990), whereas \HI should 
in general be protected from photo-dissociation by the surrounding ISM. 
Therefore CO and \HI trace different parts of the circumstellar shells.
Finally the CO line-profiles and intensities are expected to depend on 
excitation and optical depth effects much more than for H\,{\sc {i}}.

\clearpage

\begin{figure} 
\includegraphics[angle=-90,scale=.3]{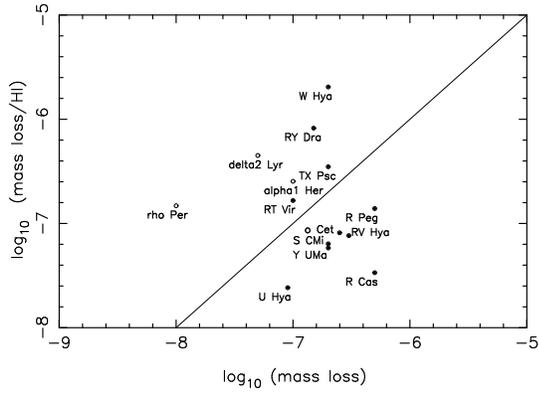}
\caption[]{Average mass loss rate estimated from \HI versus mass loss rate 
from the literature (Table~\ref{Sample_tab}). Filled circles: mass loss rates 
derived from CO rotational lines; empty circles: mass loss rates 
derived from optical data (Deutsch 1956; Sanner 1976).}
\label{masslossCO}
\end{figure}

When we compare the mass loss rates estimated from the \HI line profiles 
to those obtained from CO rotational lines, we do not find any clear 
trend (Fig.~\ref{masslossCO}). This probably indicates that the mass loss 
rate varies by an order of magnitude over timescales $\sim$ 10$^5$ years. 
Nevertheless, we note that Miras tend to lie below the one-to-one line, 
and Semi-Regulars above it. 

It thus appears that the \HI line at 21 cm and the CO rotational lines are 
complementary tracers of the circumstellar outflows and that they are equally 
needed to describe these media.

\section{DISCUSSION}\label{discussion}

\subsection{\HI Confusion}

In general our data are affected by the large scale structure of the 
galactic \HI emission. Thus the final limits of detection are not set 
by the sensitivity of the observations but by the effectiveness in 
removing the galactic emission.
When the position-switch procedure has been effective we have qualified 
the confusion as weak (Table \ref{HItab_a}). For ``medium'' and ``high'' 
confusion part of the \HI spectrum can be affected. It indicates that matter 
in a direction close to the target and at velocities in the expected range 
has been detected. In general this matter, that can be at any distance along 
the line of sight, should not be related to the stellar source.

However, as the central 
stars may have been undergoing mass loss for a long duration and/or during 
a previous stage of evolution (e.g. the RGB), in some cases it might be that 
this matter (or part of it) is physically related to the source. For instance, 
the interaction of the outflow with its surrounding leads to the 
formation of a shell made of compressed circumstellar material and swept-up 
circumstellar/interstellar matter (Sect.~\ref{lineprofiles}). The matter 
in this shell should be at velocity close to the central star velocity. 
If part of the material lost by the central star is mixed in position and in 
velocity with the surrounding ISM, we may never be able to get an exhaustive 
balance of the mass lost by the central stars. For instance, in their \HI 
study of IC~418, Taylor et al. (1989) noted that part of the detected 
hydrogen, blueshifted with respect to the PN system velocity, cannot be 
identified unambiguously as circumstellar or interstellar matter.

\subsection{Interaction with the ISM}\label{interaction_ISM}

For several sources we observe a shift between the central velocities of 
the \HI and CO lines ($\sim$ 1-3 \kms). In some cases we also observe  an 
offset of the \HI source with respect to the stellar position 
($\sim$ 1-4\arcmin). Such a kind of offset has not been observed in CO. 
Since the \HI emission profiles show evidence of a slowing-down of the stellar 
outflows by the surrounding ISM (Paper II and III, Sect.~\ref{lineprofiles}), 
and since the central stars are moving with respect to 
the ISM{\footnote{although this cannot be strictly demonstrated, 
we note that our sources are close to the Sun and have a non-zero 
\Vlsr}}, we may suspect that these velocity shifts 
and spatial offsets are due to a non-isotropic interaction between 
the outflows and the ISM. One may also consider a gradient in the ISM density 
that would induce a non-isotropic slowing-down of the circumstellar outflow.

In order to corroborate this hypothesis we have used the non-spherical model 
of \HI emission developed by Gardan et al. (2006, Paper III). As an example 
we consider a source located at 140 pc with an outflow corresponding to a mass 
loss in atomic hydrogen of 5$\times$10$^{-7}$ \Msold, and that is initially 
spherical and isotropic. The velocity is taken to decrease linearly 
from an inner boundary located at 0.1\arcmin ~(or 4$\times$10$^{-3}$ pc) from 
the central star. 
To simulate a non-isotropic interaction, we adopt a velocity law as :\\
\begin{eqnarray}
V = 10.0 - (1.43 + 0.41 cos \alpha)\times(r - 0.1)
\end{eqnarray}
with V expressed in \kms, r in \arcmin, and $\alpha$ being a polar angle. 
The slowing down of the outflow is maximum for $\alpha$ = 0$^{\circ}$ and 
minimum for $\alpha$ = 180$^{\circ}$. The geometry of the circumstellar shell 
is thus progressively becoming egg-shaped. For this test, we adopt a typical 
duration of 50$\times$10$^{3}$ years, such that the outer boundary is 
located at 5\arcmin ~(0.2 pc) for $\alpha$ = 0$^{\circ}$, with V = 1 \kms, 
and 7.2\arcmin ~($\sim$ 0.3 pc) for $\alpha$ = 180$^{\circ}$, with 
V $\sim$ 2.8 \kms. We adopted for the NRT beam 
a ``sinc'' response (see Paper III).

\begin{figure}
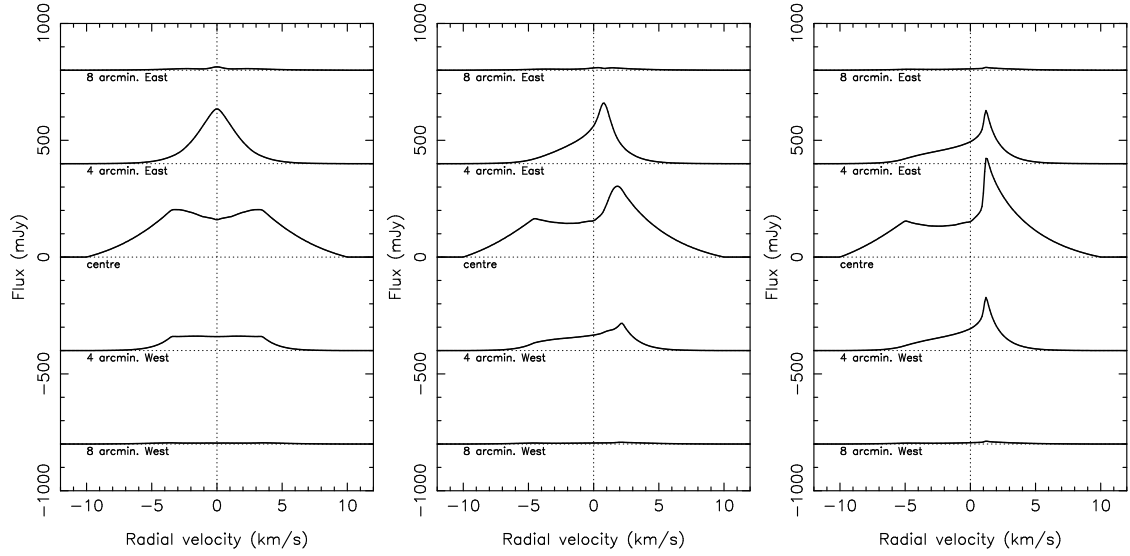

\begin{center}
\includegraphics[angle=0,scale=.30]{f26a.ps}
\includegraphics[angle=0,scale=.30]{f26b.ps}
\includegraphics[angle=0,scale=.30]{f26c.ps}
\caption[]{Simulation of the \HI emission from an egg-shaped circumstellar 
shell (see Sect.~\ref{interaction_ISM}). Left : PA = 90$^{\circ}$, 
inclination angle = 0$^{\circ}$; centre : PA = 90$^{\circ}$, 
inclination angle = $-$45$^{\circ}$; 
right : PA = 90$^{\circ}$, inclination angle = $-$90$^{\circ}$.\label{simulation}}

\end{center}
\end{figure}

The results of such a simulation are given in Fig.~\ref{simulation}. The polar 
axis of the source ($\alpha$ = 0$^{\circ}$) is first placed in the plane of 
the sky pointing East (Fig.~\ref{simulation}, left). The \HI profiles are 
symmetric and centered on the 
stellar velocity, but the emission feature is more pronounced East than West. 
If the polar axis is pointing 90$^{\circ}$ away from the observer 
(Fig.~\ref{simulation}, right), the \HI profile becomes asymmetric 
with a red-shifted peak. In that case the emission remains centered on the 
star position. Finally, if the polar axis is oriented 45$^{\circ}$ away 
from the observer and at a position angle of 90$^{\circ}$ 
(Fig.~\ref{simulation}, center), the \HI profile is asymmetric and 
the emission is stronger East of the central star.

This simulation shows that a slowing-down of the stellar outflow more 
pronounced in a given direction could produce either a shift in velocity of 
the \HI emission line, or an offset in position of this emission, or a 
combination of both. This effect may add to an intrinsic asymmetry due 
to the mass loss from the central star, as we showed for X Her (Paper III).

\subsection{Cold Atomic Hydrogen}\label{coldHI}

We have detected \HI in emission towards AFGL\,3068. However, 
we have previously reported the detection of \HI {\it in absorption} towards 
a similar source, IRC\,+10216, (Le~Bertre \& G\'erard 2001), indicating 
that the gas in the outflow cools down to a temperature below 4 K. 
Such a low temperature was predicted for high mass loss rate AGB stars
as a result of adiabatic expansion (Jura et al. 1988; Sahai 1990). 
At first sight it is puzzling that we do not observe the same 
effect in AFGL\,3068 of mass loss rate about 5 times larger. 
However the distances to these sources differ by a factor $\sim$ 8, 
so that the NRT beam scales as 0.16\,pc$\times$0.86\,pc for IRC\,+10216 
as compared to 1.3\,pc$\times$7.3\,pc for AFGL\,3068. Therefore, in the case 
of AFGL\,3068 we have explored a very external region where grain 
photo-electric 
heating might become important. For instance, in their models of IRC\,+10216, 
Crosas \& Menten (1997) and Groenewegen et al. (1998) predict that 
the temperature should increase at r $\geq$ 0.1 pc due to this effect 
(although their minima are still higher than our observational upper limit). 

The \HI detection of IRC\,+10216 in absorption and of AFGL\,3068 in emission 
might then be explained as a combined effect of beam size and distance. 
If this is correct, the \HI mass that we have derived for AFGL 3068 
may be underestimated due to absorption against the background. 

The gas temperature in the circumstellar envelopes depends strongly 
on the mass-loss rate (e.g. Jura et al. 1988). It is expected to be larger 
in the sources with low mass-loss rate ($\sim$ 10$^{-7}$ \Msold) than in those 
with high mass-loss rate ($\sim$ 10$^{-5}$ \Msold). Low mass-loss rate sources,
and in particular Semi-Regular Variables, are therefore less liable 
to contain very cold atomic hydrogen in their envelopes. On the other 
hand, this effect may concern Miras with substantial mass loss rate, 
like WX Psc or NML Tau.

\subsection{Molecular Hydrogen}\label{molhydrogen}

Hydrogen in stars with low atmospheric temperature (T$_{\star} \leq$ 2500 K) 
should be molecular (GH1983). The expected lifetime of molecular hydrogen 
in the ISM 
is $\sim$ 1000 years for a standard Interstellar Radiation Field (Le Petit 
et al. 2002). This translates to r $\sim$ 10$^{-3}$ or 10$^{-2}$ pc for 
an average expansion velocity of 1 or 10 \kms, respectively. However, 
molecular hydrogen is self-shielded (Morris \& Jura 1983) and the real extent 
of an H$_{2}$ circumstellar envelope depends on the mass loss rate and the 
velocity field. For instance GH1983 estimate that, for a mass loss rate of 
4$\times$10$^{-5}$ \Msold ~and a uniform expansion velocity of 16 \kms, 
the radius of the molecular envelope would be $\sim$ 0.2 pc. 

It is therefore important to check the presence of atomic hydrogen 
at large distance from the central star for those with low effective 
temperature. As sources with low effective temperature are also those with 
presently high mass-loss rate (e.g. WX Psc, NML Tau), 
atomic hydrogen may be detected in these sources, 
either in absorption or in emission, only at large distances from 
the central stars, and, of course, only if mass loss has been operating 
for a long enough time. 

\section{CONCLUSION}
 
We have undertaken a search for the \HI line at 21 cm 
in the direction of AGB stars and related sources. 
We have detected circumstellar \HI in a variety of sources sampling 
the post main-sequence evolution of low and intermediate mass stars.
The emissions are extended indicating shell sizes from 0.2 pc to more 
than 2 pc. The \HI line can therefore be used to trace the mass loss history 
for long periods, up to several 10$^5$ years. However, we may have missed 
atomic hydrogen at low temperature ($<$~10~K) and hydrogen locked into H$_2$.

For some sources ($\rho$ Per, $\alpha$ Her, $\delta^2$ Lyr,
U CMi) our \HI detection is the first radio detection. 
%, NGC 6369

The \HI line-profiles indicate that the outflows are slowing down, 
probably due to the interaction with surrounding matter. Furthermore, 
we observe offsets in velocity and spatial asymmetries that indicate 
a distortion of the outflows. These effects were already noted for EP Aqr 
(Paper II) and for X Her (Paper III), and seem frequent. They could be 
connected to the motion of the sources with respect to the local ISM,  
or to its inhomogeneity, as well as to the mass ejection process itself. 

%% If you wish to include an acknowledgments section in your paper,
%% separate it off from the body of the text using the \acknowledgments
%% command.

%% Included in this acknowledgments section are examples of the
%% AASTeX hypertext markup commands. Use \url without the optional [HREF]
%% argument when you want to print the url directly in the text. Otherwise,
%% use either \url or \anchor, with the HREF as the first argument and the
%% text to be printed in the second.

\acknowledgments

We are grateful to the referee for his/her detailed review that helped us
improve our paper. 
The Nan\c{c}ay Radio Observatory is the Unit\'e scientifique de Nan\c{c}ay of 
the Observatoire de Paris, associated as Unit\'e de Service et de Recherche 
(USR) No. B704 to the French Centre National de la Recherche Scientifique 
(CNRS). The Nan\c{c}ay Observatory also gratefully acknowledges the financial 
support of the Conseil R\'egional de la R\'egion Centre in France. This 
research has made use of the SIMBAD database, operated at CDS, Strasbourg, 
France and of the NASA's Astrophysics Data System.

%% To help institutions obtain information on the effectiveness of their
%% telescopes, the AAS Journals has created a group of keywords for telescope
%% facilities. A common set of keywords will make these types of searches
%% significantly easier and more accurate. In addition, they will also be
%% useful in linking papers together which utilize the same telescopes
%% within the framework of the National Virtual Observatory.
%% See the AASTeX Web site at http://www.journals.uchicago.edu/AAS/AASTeX
%% for information on obtaining the facility keywords.

%% After the acknowledgments section, use the following syntax and the
%% \facility{} macro to list the keywords of facilities used in the research
%% for the paper.  Each keyword will be checked against the master list during
%% copy editing.  Individual instruments can be provided in parentheses,
%% after the keyword, but they will not be verified.

Facilities: \facility{NRT}

%% The reference list follows the main body and any appendices.
%% Use LaTeX's thebibliography environment to mark up your reference list.
%% Note \begin{thebibliography} is followed by an empty set of
%% curly braces.  If you forget this, LaTeX will generate the error
%% "Perhaps a missing \item?".
%%
%% thebibliography produces citations in the text using \bibitem-\cite
%% cross-referencing. Each reference is preceded by a
%% \bibitem command that defines in curly braces the KEY that corresponds
%% to the KEY in the \cite commands (see the first section above).
%% Make sure that you provide a unique KEY for every \bibitem or else the
%% paper will not LaTeX. The square brackets should contain
%% the citation text that LaTeX will insert in
%% place of the \cite commands.

%% We have used macros to produce journal name abbreviations.
%% AASTeX provides a number of these for the more frequently-cited journals.
%% See the Author Guide for a list of them.

%% Note that the style of the \bibitem labels (in []) is slightly
%% different from previous examples.  The natbib system solves a host
%% of citation expression problems, but it is necessary to clearly
%% delimit the year from the author name used in the citation.
%% See the natbib documentation for more details and options.

%

\clearpage
\onecolumn  % for us
%%%%%%%%%

\begin{thebibliography}{}

\bibitem[]{1818}
Bakos, G.A., \& Tremko, J. 1991, Contr. Astron. Obs. Skalnate Pleso, 21, 99
\bibitem[]{1820}
Bergeat, J., Knapik, A., \& Rutily, B. 2001, A\&A, 369, 178 
%\bibitem[]{1822}
%Bowers, P.F., \& Knapp, G.R. 1987, ApJ, 315, 305 % alp Ori
\bibitem[]{1824} 
Bowers, P.F., \& Knapp, G.R. 1988, ApJ, 332, 299 (BK1988) % Mira
\bibitem[]{1826}
Crosas, M., \& Menten, K.M. 1997, ApJ, 483, 913
\bibitem[]{1828}
Deutsch, A.J. 1956, ApJ, 123, 210
\bibitem[]{1830}
Dumm, T., \& Schild, H. 1998, New Astronomy, 3, 137
\bibitem[]{1832}
Dyck, H.M., Lockwood, G.W., \& Capps, R.W. 1974, ApJ, 189, 89
%\bibitem[]{1834}
%Dyck, H.M., van Belle, G.T., \& Thompson R.R. 1998, AJ, 116, 981
\bibitem[]{1836}
Etoka, S., Le Squeren, A.M., \& G\'erard, E. 2003, A\&A, 403, L51
\bibitem[]{}
Feast, M.W. 1996, MNRAS, 278, 11
\bibitem[]{}
Garay, G., Gathier, R., Rodr\'iguez, L.F. 1989, A\&A, 215, 101
\bibitem[]{1840}
Gardan, E., G\'erard, E., \& Le~Bertre, T. 2006, MNRAS, 365, 245 (Paper III)
\bibitem[]{1842}
Gehrz, R.D., Hackwell, J.A., \& Briotta, D. 1978, ApJ, 221, L23
\bibitem[]{1844}
G\'erard, E., \& Le~Bertre, T. 2003, A\&A, 397, L17 (Paper I)
\bibitem[]{1846}
Glassgold, A.E., \& Huggins, P.J. 1983, MNRAS, 203, 517 (GH1983)
\bibitem[]{1848}
Gonz\'alez Delgado, D., Olofsson, H., Kerschbaum, F., Sch{\" o}ier, F.L., 
Lindqvist, M., \& Groenewegen M.A.T. 2003, A\&A, 411, 123
\bibitem[]{1851}
Groenewegen, M.A.T., van der Veen, W.E.C.J., \& Matthews, H.E. 1998, 
A\&A, 338, 491
\bibitem[]{}
Groenewegen, M.A.T., \& Whitelock, P.A. 1996, MNRAS, 281, 1347
\bibitem[]{1854}
Guilain, C., \& Mauron, N. 1996, A\&A, 314, 585
\bibitem[]{1856}
Haniff, C.A., Scholz, M., \& Tuthill, P.G. 1995, MNRAS, 276, 640
\bibitem[]{1858}
Hartmann, D., \& Burton, W.B. 1997, Atlas of Galactic Neutral 
Hydrogen, Cambridge University Press
\bibitem[]{}
Hashimoto, O., \& Izumiura, H. 2000, Advances in Space Research, 25, 2197
\bibitem[]{1861}
Heras, A.M., Shipman, R.F., Price, S.D., et al. 2002, A\&A 394, 539 
\bibitem[]{1863}
Heske, A. 1990, A\&A, 229, 494
\bibitem[]{1865}
Heske, A., te Lintel Hekkert, P., \& Maloney, P.R. 1989, A\&A, 218, L5
\bibitem[]{1867}
Hony, S., Waters, L.B.F.M., \& Tielens, A.G.G.M. 2001, A\&A, 378, L41
\bibitem[]{1869}
Huggins, P.J., Cox, P., Forveille, T., Bachiller, R., \& Young, K. 1999, 
IAU~S191, 425    % NGC 7293 
\bibitem[]{1872}
Izumiura, H., \& Hashimoto, O. 1999, IAU S191, 401
\bibitem[]{1874}
Jones, B., Merrill, K.M., Puetter, R.C., \& Willner, S.P. 1978, AJ, 83, 1437
\bibitem[]{1876}
Josselin, E., Mauron, N., Planesas, P., \& Bachiller, R. 2000, A\&A, 362, 255
%\bibitem[]{1878}
%Kahane, C., \& Jura, M. 1996, A\&A, 310, 952
\bibitem[]{1880}
Jura, M., Kahane, C., \& Omont, A. 1988, A\&A, 201, 80
\bibitem[]{1882}
Justtanont, K., Bergman, P., Larsson, B., et al. 2005, A\&A, 439, 627
\bibitem[]{1884}
Knapp, G.R., Crosas, M., Young, K., \& Ivezi\'{c}, \v{Z}. 2000, ApJ, 534, 324
\bibitem[]{1886}
Knapp, G.R., \& Morris, M. 1985, ApJ, 292, 640
\bibitem[]{1888}
Knapp, G.R., Young, K., Lee, E., \& Jorissen, A. 1998, ApJS, 117, 209
\bibitem[]{1890}
Lamers, H.J.G.L.M., \& Cassinelli, J.P. 1999, Introduction to Stellar Winds, 
Cambridge University Press
\bibitem[]{1893}
Le~Bertre, T. 1992, A\&AS, 94, 377
\bibitem[]{1895}
Le~Bertre, T. 1993, A\&AS, 97, 729
\bibitem[]{1897}
Le~Bertre, T. 1997, A\&A, 324, 1059
\bibitem[]{1899}
Le~Bertre, T., \& G\'erard, E. 2001, A\&A, 378, L29
\bibitem[]{1901}
Le~Bertre, T., \& G\'erard, E. 2004, A\&A, 419, 549 (Paper II)
\bibitem[]{1903}
Le~Bertre, T., Gougeon, S., \& Le Sidaner, P. 1995, A\&A, 299, 791
\bibitem[]{1905}
Le~Bertre, T., \& Winters J.M. 1998, A\&A, 334, 173
\bibitem[]{1907}
Le~Petit, F., Roueff, E., \& Le Bourlot, J. 2002, A\&A, 390, 369
\bibitem[]{1909}
Le~Sidaner, P., \& Le~Bertre, T. 1996, A\&A, 314, 896
\bibitem[]{1911}
Loup, C., Forveille, T., Omont, A., \& Paul, J.F. 1993, A\&AS, 99, 291
\bibitem[]{1913}
Mamon, G.A., Glassgold, A.E., \& Huggins, P.J. 1990, ApJ, 328, 797
\bibitem[]{1915}
Mauron, N., \& Caux, E. 1992, A\&A, 265, 711
\bibitem[]{1917}
Mauron, N., \& Guilain, C. 1995, A\&A, 298, 869
\bibitem[]{1919}
Mauron, N., \& Huggins, P.J. 2006, A\&A, 452, 257
\bibitem[]{1921}
Meatheringham, S.J., Wood, P.R., \& Faulkner, D.J. 1988, ApJ, 334, 862
\bibitem[]{1923}
Monteiro, H., Schwarz, H.E., Gruenwald, R., \& Heathcote, S. 2004, 
ApJ, 609, 194
\bibitem[]{1926}
Morris, M., \& Jura, M. 1983, ApJ, 264, 546
\bibitem[]{1928}
Neri, R., Kahane, C., Lucas, R., Bujarrabal, V., \& Loup, C. 1998, A\&AS, 
130, 1
\bibitem[]{1931}
Nyman, L.-\AA., Booth, R.S., Carlstr{\" o}m, U., et al. 1992, A\&AS, 93, 121
\bibitem[]{1933}
Olofsson, H., Gonz{\'a}lez~Delgado, D., Kerschbaum, F., \& Sch{\"o}ier, F.L. 
2002, A\&A, 391, 1053
\bibitem[]{1936}
Perrin, G., Coud\'e du Foresto, V., Ridgway, S.T., et al. 1998, A\&A, 331, 619
\bibitem[]{1938}
Perrin, G., Ridgway, S.T., Coud\'e du Foresto, V., et al. 2004a, A\&A, 418, 675
\bibitem[]{1940}
Perrin, G., Ridgway, S.T., Mennesson, B., et al. 2004b, A\&A, 426, 279
\bibitem[]{1942}
Reich, P., \& Reich, W. 1986, A\&AS, 63, 205 
\bibitem[]{1944}
Reid, M.J., \& Menten, K.M. 1997, ApJ, 476, 327
%\bibitem[]{1946}
%Reimers, D. 1977, A\&A, 61, 217 (1978, A\&A, 67, 161)
\bibitem[]{1948}
Reimers, D., \& Cassatella, A. 1985, ApJ, 297, 275
\bibitem[]{1950}
Ridgway, S.T., Joyce, R.R., White, N.M., \& Wing, R.F. 1980, ApJ, 235, 126
 \bibitem[]{1952}
Rodr{\'\i}guez, L.F., Goss, W.M., \& Williams, R. 2002, ApJ, 574, 179 (R2002)
% NGC 7293
\bibitem[]{1955}
Ryde, N., \& Sch\"oier, F.L. 2001, ApJ, 547, 384
\bibitem[]{1957}
Sahai, R. 1990, ApJ, 362, 652
%\bibitem[]{1959}
%Sahai, R., Morris, M., Knapp, G.R., Young, K., \& Barnbaum, C. 2003, 
%Nature, 426, 261
\bibitem[]{1962}
Sanner, F. 1976, ApJS, 32, 115
\bibitem[]{1964}
Sivagnanam, P., Le Squeren, A.M., Foy, F., \& Tran Minh, F. 1989, A\&A, 
211, 341
\bibitem[]{1967}
Speck, A.K., Barlow, M.J., Sylvester, R.J., \& Hofmeister, A.M. 2000, 
A\&AS, 146, 437 
\bibitem[]{1970}
Speck, A.K., Meixner, M., Fong, D., McCullough, P.R., Moser, D.E., \& 
Ueta, T. 2002, AJ, 123, 346
\bibitem[]{1973}
Sudol, J.J., Benson, J.A., Dyck, H.M., \& Scholz, M. 2002, AJ, 124, 3370
\bibitem[]{1975}
Taylor, A.R., Gussie, G.T., \& Goss, W.M. 1989, ApJ, 340, 932
\bibitem[]{1977}
Taylor, A.R., Gussie, G.T., \& Pottasch, S.R. 1990, ApJ, 351, 515
\bibitem[]{1979}
van Belle, G.T., Dyck, H.M., Benson, J.A., \& Lacasse, M.G. 1996, AJ, 112, 2147
\bibitem[]{1981}
van Driel W., Pezzani J., \& G\'erard E. 1996, in ``High Sensitivity Radio 
Astronomy'', N. Jackson \& R.J. Davis (eds.), Cambridge Univ. Press, p.~229
\bibitem[]{1984}
Volk, K., \& Cohen, M. 1989, AJ, 98, 931
\bibitem[]{1986}
Waters, L.B.F.M., Loup, C., Kester, D.J.M., Bontekoe, Tj.R., \& de Jong, T. 
1994, A\&A, 281, L1
\bibitem[]{1989}
Weiner, J. 2004, ApJ, 611, L37
\bibitem[]{1991}
Winters, J.M., Fleischer, A.J., Le~Bertre, T., \& Sedlmayr, E. 1997, A\&A, 
326, 305
\bibitem[]{1994}
Winters, J.M., Le~Bertre, T., Jeong, K.S., Nyman, L.-\AA., \& Epchtein, N.   
2003, A\&A, 409, 715
\bibitem[]{1997}
Woitke, P., Helling, C., Winters, J.M., \& Jeong, K.S. 1999, A\&A, 348, L17
\bibitem[]{1999}
Young, K. 1995, ApJ, 445, 872
\bibitem[]{2001}
Young, K., Cox, P., Huggins, P.J., Forveille, T., \& Bachiller, R. 1997, ApJ, 
482, L101
\bibitem[]{2004}
Young, K., Cox, P., Huggins, P.J., Forveille, T., \& Bachiller, R. 1999, ApJ, 
522, 387
\bibitem[]{2007}
Young, K., Phillips, T.G., \& Knapp, G.R. 1993a, ApJS, 86, 517
\bibitem[]{2009}
Young, K., Phillips, T.G., \& Knapp, G.R. 1993b, ApJ, 409, 725
\bibitem[]{2011}
Zuckerman, B., Terzian, Y., \& Silverglate, P. 1980, ApJ, 241, 1014
\end{thebibliography}
\end{document}